\documentclass[aps,prl,floats,showpacs]{revtex4}
\usepackage{graphicx}

\begin{document}

\def\nuc#1#2{${}^{#1}$#2}
\def\mee{$\langle m_{\beta\beta} \rangle$}
\def\mnu{$\langle m_{\nu} \rangle$}
\def\gnu{$\langle g_{\nu,\chi}\rangle$}
\def\mmod{$\| \langle m_{ee} \rangle \|$}
\def\mb{$\langle m_{\beta} \rangle$}
\def\BBz{$0\nu\beta\beta$}
\def\BBm{$0\nu\chi\beta\beta$}
\def\BBt{$2\nu\beta\beta$}
\def\BB{$\beta\beta$}
\def\bb{$\beta\beta$}
\def\bdec{\mbox{$\beta$ decay}}
\def\Mz{$|M_{0\nu}|$}
\def\ezero{$E_0$}
\def\Mt{$|M_{2\nu}|$}
\def\Tz{$T^{0\nu}_{1/2}$}
\def\Tt{$T^{2\nu}_{1/2}$}
\def\Tc{$T^{0\nu\,\chi}_{1/2}$}
\def\ms{$\Delta m_{\rm sol}^{2}$}
\def\ma{$\Delta m_{\rm atm}^{2}$}
\def\ts{$\theta_{\rm sol}$}
\def\ta{$\theta_{\rm atm}$}
\def\tot{$\theta_{13}$}
\def\ep{\mbox{e$^{+}$}}
\def\el{\mbox{e$^{-}$}}
\def\pos{\mbox{e$^{+}$}}
\def\belec{\mbox{$\beta$-electron}}
\def\bspec{\mbox{$\beta$ spectrum}}
\def\nue{\mbox{$\nu_{e}$}}
\def\numu{\mbox{$\nu_{\mu}$}}
\def\numub{\mbox{$\bar{\nu}_{\mu}$}}
\def\nub{\mbox{$\bar{\nu}$}}
\def\nueb{\mbox{$\bar{\nu}_{e}$}}
\def\nubx{\mbox{$\bar{\nu}_{x}$}}
\def\nutau{\mbox{$\nu_\tau$}}
\def\nui{\mbox{$\nu_1$}}
\def\nuii{\mbox{$\nu_2$}}
\def\nuiii{\mbox{$\nu_3$}}

\def\etal{{\it et al.}}
\def\be{\begin{equation}}
\def\ee{\end{equation}}

\def\Jo#1#2#3#4{{\it #1} {\bf #2}, #3 (#4)}
\def\bi#1{#1}
\def\NPBP{{Nucl. Phys.} {\bf B} (Proc. Suppl.)}
\def\NPB{{Nucl. Phys.} {\bf B}}
\def\PLB{{Phys. Lett.}  {\bf B}}
\def\PRL{Phys. Rev. Lett.}
\def\PRD{{Phys. Rev.} {\bf D}}
\def\EPC{{Eur. Phys. J.} {\bf C}}
\def\EPA{{Eur. Phys. J.} {\bf A}}
\def\ZPC{{Z. Phys.} {\bf C}}
\def\IJMP{Int. J. of Mod. Phys. {\bf A}}
\def\JHEP{JHEP}
\def\PTP{Prog. Theo. Phys.}
\def\MPLA{{Mod. Phys. Lett} {\bf A}}
\def\PPNP{Prog. Part. Nucl. Phys.}
\def\NIMA{{Nucl. Instrum. Methods} A}




\title{Neutrinoless double beta decay and direct searches for neutrino mass
\footnote{Full texts of the report of the working group. For the 
summary report of the APS Multidivisional Neutrino Study, 
`The Neutrino Matrix', see physics/0411216}.  } 

\author{Craig Aalseth$^1$,
Henning Back$^2$,
Loretta Dauwe$^3$,
David Dean$^4$,
Guido Drexlin$^5$,
Yuri Efremenko$^6$,
Hiro Ejiri$^7$,
Steven Elliott$^8$,
Jon Engel$^9$,
Brian Fujikawa$^{10}$,
Reyco Henning$^{10}$,
G. W. Hoffmann$^{11}$,
Karol Lang$^{11}$,
Kevin Lesko$^{10}$,
Tadafumi Kishimoto$^7$,
Harry Miley$^1$,
Rick Norman$^{12}$,
Silvia Pascoli$^{13}$,
Serguey Petcov$^{14}$,
Andreas Piepke$^{15}$,
Werner Rodejohann$^{16}$,
David Saltzberg$^{13}$,
Sean Sutton$^{17}$,
Petr Vogel$^{18}$,
Ray Warner$^1$,
John Wilkerson$^{19}$,
and Lincoln Wolfenstein$^{20}$.}

\affiliation{
$^1$Pacific Northwest National Laboratory \\
$^2$Virgina Tech \\
$^3$University of Michigan \\
$^4$Oak Ridge National Laboratory \\
$^5$Forschungszentrum Karlsruhe, Institut fur Kernphysik \\
$^6$University of Tennesee \\
$^7$Osaka University \\
$^8$Los Alamos National Laboratory \\
$^9$University of North Carolina \\
$^{10}$Lawrence Berkeley National Laboratory \\
$^{11}$University of Texas at Austin \\
$^{12}$Lawrence Livermore National Laboratory \\
$^{13}$University of California, Los Angeles \\
$^{14}$SISSA Sezione di Trieste \\
$^{15}$University of Alabama \\
$^{16}$SISSA and INFN Sezione di Trieste \\
$^{17}$Mount Holyoke College \\
$^{18}$California Institute of Technology \\
$^{19}$University of Washington \\
$^{20}$Carnegie Mellon University
}

\date{\today}

\begin{abstract}
 Study of the neutrinoless double beta decay and searches for the manifestation
of the neutrino mass in ordinary beta decay are the main sources of information
about the absolute neutrino mass scale, and the only practical source
of information about the charge conjugation properties of the neutrinos.
Thus, these studies have a unique role in the plans for better understanding
of the whole fast expanding field of neutrino physics.

\end{abstract}


\maketitle

\section{I. Executive Summary}


\medskip
The physics addressed by this research program seeks to answer 
many of the Study's questions:  
\begin{enumerate}
\item Are neutrinos their own anti-particles?
\item What are the masses of the neutrinos?
\item Do neutrinos violate the symmetry CP?
\item Are neutrinos the key to the understanding of the matter-antimatter asymmetry of the 
Universe?
\item What do neutrinos have to tell us about the intriguing proposals for new models of physics?
\end{enumerate}

Only the research covered within this working group can 
answer the first  and second of these fundamental questions.
Among the ways to measure the neutrino mass, 
three are notable because they are
especially sensitive: double-beta decay, tritium beta decay, 
and cosmology. Consequently, we have focused our report 
and recommendations on them. 

$\bullet$ Observation of the  neutrinoless double-beta decay ($0\nu\beta\beta$) 
would prove that the total lepton number is not conserved
and would establish a non vanishing
neutrino mass of Majorana nature. 
In other words, observation of the $0\nu\beta\beta$ decay, independent
of its rate, would show that neutrinos, unlike all the other
constituents of matter, are their own antiparticles.
There is no other realistic way
to determine the nature - Dirac or Majorana, of massive neutrinos. 
This would be a discovery of major importance, with impact not only on
this fundamental question, but also on the determination of the absolute
neutrino mass scale, and on the pattern of neutrino masses, and possibly 
on the problem of CP violation in the lepton sector,
associated with Majorana neutrinos. 
There is a consensus on this basic point
which we translate into the recommendations
how to proceed with experiments dedicated to the
search of the  $0\nu\beta\beta$ decay, and how to fund them.

$\bullet$ To reach our conclusion, we  have to consider past achievements,
the size of previous experiments, and the existing proposals. 
There is a considerable community
of physicists worldwide as well as in the US interested in pursuing the
search for the  $0\nu\beta\beta$ decay. Past experiments were
of relatively modest size. Clearly, the scope
of future experiments should be considerably larger, and will require
advances in experimental techniques, larger collaborations 
and additional funding.
In terms of $\langle m_{\beta\beta} \rangle$, the effective neutrino Majorana
mass that can be extracted 
from the observed  $0\nu\beta\beta$ decay rate, there are three ranges
of increasing sensitivity, related to known neutrino-mass scales 
of neutrino oscillations.

$\bullet$ The $\sim$100-500 meV  $\langle m_{\beta\beta} \rangle$
range corresponds to the quasi-degenerate
spectrum of neutrino masses. The motivation for reaching 
this scale has been strengthened by the recent
claim of an observation of $0\nu\beta\beta$ decay in $^{76}$Ge;
a claim that obviously requires further investigation. To reach this scale
and perform reliable measurements,
the size of the experiment should be approximately 
200 kg of the decaying isotope, 
with a corresponding reduction of the background.

This quasi-degenerate scale is achievable in the relatively
near term, $\sim$ 3-5 years.
Several groups with considerable US participation have well established
plans to build $\sim$ 200-kg devices that could scale straight-forwardly
to 1 ton (Majorana using $^{76}$Ge, Cuore using  $^{130}$Te, 
and EXO using  $^{136}$Xe). There are also other proposed experiments worldwide which
offer to study a number of other isotopes and
could reach similar sensitivity after further R\&D.
Several among them ({\it e.g.} Super-NEMO, MOON) have US participation.

By making measurements in several
nuclei the uncertainty arising from the nuclear
matrix elements would be reduced. The development of different
detection techniques, and measurements in several nuclei, 
is invaluable for establishing the existence
(or lack thereof) of the  $0\nu\beta\beta$ decay at this
effective neutrino mass range.       

$\bullet$ The $\sim$20-55 meV range arises from the atmospheric
neutrino oscillation results.  Observation of 
$\langle m_{\beta\beta} \rangle$ at this mass scale 
would imply the inverted
neutrino mass hierarchy or the normal-hierarchy 
$\nu$ mass spectrum very near the quasi-degenerate
region.
If either this or the quasi-degenerate spectrum is established,
it would be invaluable not only for the understanding of the 
origin of neutrino mass, but also as input 
to the overall neutrino physics program 
(long baseline oscillations, search for
CP violations, search for neutrino mass in tritium beta decay
and astrophysics/cosmology, etc.)  

To study the 20-50 meV mass range will require about 1
ton of the isotope mass, a  challenge of its own.
Given the importance, and the points discussed above, more 
than one experiment of that size is desirable.

$\bullet$ The $\sim$2-5 meV range arises from the solar
neutrino oscillation results and will almost certainly lead
to the  $0\nu\beta\beta$ decay, provided neutrinos are
Majorana particles. 
To reach this goal will require $\sim$100 tons of the 
decaying isotope,
and no current technique provides such a leap in sensitivity. 

$\bullet$ The qualitative physics results that arise from an observation
of $0\nu\beta\beta$ decay are profound. Hence, the program described above
is vital and fundamentally important
even if the resulting $\langle m_{\beta\beta} \rangle$ would be 
rather uncertain in value.
However,  by making measurements in several
nuclei the uncertainty arising from the nuclear
matrix elements would be reduced.

$\bullet$ Unlike double-beta decay, beta-decay endpoint measurements  search
for a kinematic effect due to neutrino mass and  thus are 
"direct searches" for neutrino mass. This technique,
which is essentially free of theoretical assumptions
about neutrino properties, is not just complementary. In fact,
both types of measurements will be required to fully untangle 
the nature of the neutrino mass.
Excitingly, a very large new beta spectrometer is being built in Germany. 
This KATRIN experiment
has a design sensitivity approaching 200 meV. 
If the neutrino masses are quasi-degenerate, as would
be the case if the recent double-beta decay claim proves true, 
KATRIN will see the effect. In this case the $0\nu\beta\beta$-decay experiments 
can provide, in principle, unique information
about CP-violation in the lepton sector, associated with Majorana 
neutrinos.

$\bullet$ Cosmology can also provide crucial information on the sum of the
neutrino masses. This topic is summarized in a different section of the report, but it 
should be mentioned here that the next generation of measurements 
hope to be able to observe a sum of neutrino masses as small as 40 meV. 
We would like to emphasize the complementarity of the three approaches, $0\nu\beta\beta$ ,
$\beta$ decay,
and cosmology. 

\medskip

{\bf Recommendations:}

We conclude that such a double-beta-decay program can be summarized as having three
components and our recommendations
can be summarized as follows:
\begin{enumerate}
\item A substantial number (preferably more
than two) of 200-kg scale experiments (providing the capability to
make a precision measurement
at the quasi-degenerate mass scale) with large US participation
should be supported as soon as possible. \\
$\circ$ Each such experiment will cost approximately \$10M-\$20M and
take 3-5 years to implement.
\item Concurrently, the development
toward $\sim$1-ton experiments ({\it i.e.} sensitive to $\sqrt{\Delta
m_{\rm atm}^{2}}$)
should be supported, primarily as expansions of the 200-kg experiments.
The corresponding plans for the procurement
of the enriched isotopes, as well as for the development of a suitable
underground facility, should be carried out. The US funding agencies
should set up in a timely manner a mechanism to review and compare
the various proposals for such experiments which span
research supported by the High Energy and Nuclear Physics offices of DOE
as well as by NSF. \\
$\circ$ Each such experiment will cost approximately \$50M-\$100M and
take 5-10 years to implement.
\item A diverse R\&D program developing
additional techniques should be supported.\\
{\it The total cost of this described program will be approximately
\$250M over a 10 year period.}
\end{enumerate}

$\bullet$ In addition to double-beta decay, other techniques for
exploring the neutrino mass
need to be pursued also. We summarize these recommendations as follows.
\begin{enumerate}
\item Although the KATRIN is predominately a European effort, there
is significant US participation.
The design and construction of this experiment is proceeding well and
the program should
continue to be strongly supported.
\item Research and development of other techniques for observing the
neutrino mass kinematically
should be encouraged.
\end{enumerate}

\newpage
\section{II. Introduction}
The standard model of electroweak interactions,
developed in the late 1960's, incorporated  neutrinos as left-handed 
massless partners of the charged leptons.  
The discovery of the third generation of quarks and 
leptons completed  the model, and made it possible, 
in addition, to incorporate also a description of 
CP violation. Later efforts to unify the strong and
electroweak interactions led to the development
of Grand Unified Theories which provided a natural framework 
for neutrino masses, and motivated many experiments in the field.
Studies of $e^+e^-$ annihilation at the $Z$-resonance
peak have determined the invisible width of the $Z$ boson, caused by
its decay into unobservable channels. Interpreting this width
as a measure of the number of neutrino active flavors, one can,
quite confidently,
conclude that there are just three active neutrinos with masses of
less than $M_Z/2$.

In parallel, the understanding of  big-bang nucleosynthesis  
and the discovery of the cosmic microwave background illustrated
the important role of neutrinos in the history of the
early universe. Those developments also led to the possibility
that neutrinos, with small but finite
mass, could explain the existence of dark matter.
Although it now appears that neutrinos are not a dominant 
source of dark matter in the universe, the
experimental evidence obtained in the last decade
for finite neutrino masses and mixing between generations is strong
and compelling. Those discoveries, the first
evidence for `physics beyond the Standard Model' gives us
an intriguing glimpse into the fundamental source of particle mass
and the role of flavor 
in the scheme of particles and their interactions.
The scale of neutrino mass differences motivates 
experimental searches for the neutrinoless double beta decay 
and end-point anomalies in beta decay described in this report.

Despite the recent triumphs of neutrino physics, several fundamental
questions remain to be answered to advance the field itself and its
impact in general on the whole 
particle and nuclear physics, as well as astrophysics and
cosmology. The studies of neutrinoless double beta decay 
and end-point anomalies in beta decay, in particular, are essential
and unique in their potential to answer the first two of them
and plays an important and equally unique role in the remaining ones: 
\begin{itemize}
\item  Are neutrinos their own anti-particles?
\item What are the masses of the neutrinos?
\item Do neutrinos violate the symmetry CP?
\item Are neutrinos the key to the understanding of the matter-antimatter asymmetry of the
Universe?
\item What do neutrinos have to tell us about the intriguing proposals for new models of physics?
\end{itemize}

The present report is structured as follows: In this introductory section
we provide the ``Goal of the field'' statement first in which we stress
the fundamental importance of the distinction between the Dirac and Majorana
neutrinos, and its relation to the existence of the neutrinoless
double beta decay ($0\nu \beta\beta$). 

In the next section 
``$0\nu \beta\beta$ and $\beta$ decays and Oscillations'' 
we briefly summarize the
status of the neutrino oscillation studies and the values of the corresponding
mass differences $\Delta m_{ij}^2$ and mixing angles. Next we
discuss the relations and constraints provided by the
results of oscillation studies and the neutrino mass parameters
extracted from $0\nu \beta\beta$ and $\beta$ decay experiments. 
We also show how this
research fits into the larger picture of the whole neutrino
field described in the other reports of this APS study.
Moreover, we stress the importance of Majorana neutrinos for leptogenesis,
i.e. for the explanation of the (tiny) baryon/photon ratio
and the present excess of baryons over antibaryons.

For $0\nu \beta\beta$, the process that is observable only in heavy nuclei,
the understanding of the nuclear structure plays an essential role
in extracting the neutrino effective mass from the observed rate.
We discuss the nuclear structure aspects in the section titled
``Nuclear Structure Issues'' and comment
on the existing uncertainties as well as on the prospects of reducing
them.

The rest of the report deals with the experiments. In the section
titled ``Experimental Prospects for \BB'' we summarize the situation in
experimental \BB\ and briefly describe the numerous proposals. A similar
treatment for $\beta$ decay is discussed in ``Experimental Prospects for $\beta$ Decay''.
The relationship between cosmology and neutrino mass is described
in the final section ``Cosmology and Neutrino Mass". We end with a conclusion section.

Throughout the report we often use results reported in earlier
reviews by some of us \cite{ELL02,McV04,ELL04,PasPet03}. 

At present, we do not know the absolute scale of the neutrino mass. There is
an upper limit, of $\sim$ few eV from combining the limits from the tritium
$\beta$ decay with the $\Delta m^2$ values from the oscillation studies.
For some, but not all, of the neutrinos there is also a lower limit, simply
$\sqrt{\Delta m^2}$. These limits show that neutrinos, while massive,
are very much lighter than the other fundamental constituents of matter,
the charged leptons and quarks. While we do not understand the mass values
of any fermion, the huge difference in masses of neutrinos and all charged
fermions clearly requires an explanation. The usual one, like the see-saw
mechanism, ties the neutrino mass with some very high mass scale.
It also suggests that neutrinos, unlike all other fermions, are
Majorana particles, i.e. they are their own antiparticles.

The research discussed here, if successfull, would show whether these ideas
are true or not. As stated already in the executive summary, 
observation of the  $0\nu\beta\beta$ decay
would prove that the total lepton number is not conserved
and would establish a nonvanishing
neutrino mass of Majorana nature.
In other words, observation of the \BBz\ decay, independently
of its rate, would show that neutrinos, unlike all the other
constituents of matter, are their own antiparticles.
There is no other realistic way
to determine the nature - Dirac or Majorana, of massive neutrinos.
This would be a discovery of major importance, comparable to
the already discovered oscillations of atmospheric, solar and reactor
neutrinos, and as important as a discovery of
$CP$ violation involving neutrinos. It would have impact not only on
this fundamental question, but also on the determination of the absolute
neutrino mass scale, and on the pattern of neutrino masses, and possibly
on the problem of CP-violation in the lepton sector,
associated with Majorana neutrinos.

At the same time,  beta-decay endpoint measurements  search
for a kinematic effect due to neutrino mass and therefore are frequently
referred to
as "direct searches" for neutrino mass. This technique,
which is essentially free of theoretical assumptions
about neutrino properties, is not just complementary
to the search of \BBz\ decay. In fact,
both types of measurements will be required to fully untangle
the nature of the neutrino mass.

The following sections describe the status of the field, and plans
for further experiments. Determining the absolute neutrino mass
scale, and finding whether neutrinos are indeed Majorana particles and
thus that the lepton number is not conserved, would represent a major
advance in our understanding of particle physics.

\section{\BBz\ and $\beta$ decay and Oscillations}
\subsection{Status of oscillation searches}

As is well known, the concept of neutrino oscillations is based on the
assumption that the neutrinos of definite flavor 
($\nu_e, \nu_{\mu}, \nu_{\tau}$)
are not necessarily states of a definite mass $\nu_1, \nu_2, \nu_3 ...$.  
Instead, they are 
generally coherent superpositions of such states,
\begin{equation}
| \nu_{\ell} \rangle = \sum_i U_{\ell i} | \nu_i \rangle ~.
\label{e:super}
\end{equation} 
When the standard model is extended to include neutrino mass, the
mixing matrix $U$ is unitary. As a consequence 
the neutrino flavor is no longer a conserved quantity and
for neutrinos propagating in vacuum
the amplitude of the process $\nu_{\ell} \rightarrow \nu_{\ell'}$ is
\begin{equation}
A(\nu_{\ell} \rightarrow \nu_{\ell'}) = 
\sum_i U_{\ell i}e^{-i\frac{m_i^2 L}{2E}}
U_{\ell'i}^* ~,
\label{e:ampl}
\end{equation}
The probability of the flavor change for $\ell \ne \ell'$
is  the square of this amplitude, 
$P(\nu_{\ell} \rightarrow \nu_{\ell'}) = |A(\nu_{\ell} \rightarrow \nu_{\ell'})|^2$.
It is obvious that due to the unitarity of $U$ there is no flavor
change if all masses vanish or are exactly degenerate.
The idea of oscillations was discussed early on
by Pontecorvo \cite{Pont1,Pont2}
and by Maki, Nakagawa and Sakata \cite{MNS}. Hence, the mixing matrix $U$
is often associated with these names and the notation $U_{MNS}$ or $U_{PMNS}$
is used.

The formula for the probability is  particularly simple when only two
neutrino flavors, $\nu_{\ell}$ and  $\nu_{\ell'}$, mix appreciably,
since only one mixing angle and two neutrino masses $m_i, m_j$ are then relevant,
\begin{equation}
P(\nu_{\ell} \rightarrow \nu_{\ell' \ne \ell}) =
\sin^22\theta \sin^2 \left[ 1.27 |\Delta m_{ji}^2| ({\rm eV^2})
\frac{L({\rm km})}{E_{\nu}({\rm GeV})} \right] ~,
\label{e:osc2}
\end{equation}
where the appropriate factors of $\hbar$ and $c$ were included. 
Here $\Delta m^2_{ji} \equiv m_j^2 - m_i^2$ is the mass squared difference.
Due to the large difference in the two observed $\Delta m^2$ values,
this simple formula adequately describes most of the experiments
as of now.

In general,
the mixing matrix of 3 neutrinos is parametrized by three angles,
conventionally denoted as
$\theta_{12}, \theta_{13},  \theta_{23}$,  one $CP$ violating phase $\delta$
and two Majorana phases $\alpha_1, \alpha_2$ \cite{BHP80,SV80,Doi81}. 
Using $c$ for the cosine and $s$ for the sine, the mixing matrix $U$ 
is parametrized as

\begin{eqnarray}
\hspace{-1.6cm} \left( \begin{array}{c}
\nu_e \\ \nu_{\mu} \\ \nu_{\tau}
\end{array} \right)
  =
 \left( \begin{array}{ccc}
c_{12}c_{13} & s_{12}c_{13} & s_{13} \\
-s_{12}c_{23}-c_{12}s_{23}s_{13}e^{i\delta} &
c_{12}c_{23}-s_{12}s_{23}s_{13}e^{i\delta} &
s_{23}c_{13} e^{i\delta} \\
s_{12}s_{23}-c_{12}c_{23}s_{13}e^{i\delta} &
-c_{12}s_{23}-s_{12}c_{23}s_{13}e^{i\delta} &
c_{23}c_{13} e^{i\delta}
\end{array} \right)
\left( \begin{array}{r}
e^{i\alpha_1/2}~\nu_1 \\ e^{i\alpha_2/2}~\nu_2 \\ \nu_3
\end{array} \right)  ~.
\label{e:u3}
\end{eqnarray}
The three neutrino masses $m_i$ should be added to the parameter
set that describes the matrix (\ref{e:u3}), representing therefore
nine unknown parameters altogether.

The evidence for
oscillations of solar ($\nu_e$) and
atmospheric ($\nu_{\mu}$ and $\bar{\nu}_{\mu}$)
neutrinos is compelling and generally accepted.

Evidence for oscillations of the solar
$\nu_e$
have been reported first by
the pioneering Davis et al. (Homestake)
experiment \cite{Davis68}. It has
been confirmed and reinforced
later by Kamiokande, SAGE, GALLEX/GNO
and Super-Kamiokande experiments
\cite{Cl98,SKsol}.

The SNO solar neutrino experiment \cite{SNO,SNO3}, in which it is
possible to separately determine the flux of $\nu_e$
neutrinos reaching the detector (through the charged
current reactions) and the flux of all
active neutrinos (through the neutral current reactions),
made the conclusion that solar neutrinos oscillate,
inescapable.

Independently, KamLAND  reactor antineutrino experiment has
shown that $\bar{\nu}_e$ neutrinos oscillate as well
\cite{KamLAND1,KamLAND2}. Moreover, the oscillation
parameters extracted from that experiment agree perfectly
with those from the solar $\nu_e$ experiments. This agreement, 
expected by the CPT-invariance, shows that the formalism
of oscillations, including the matter effects, is well
understood. Based on the combined analysis of these data
the parameters $\Delta m^2_{21}$ (including its positive sign)
and $\theta_{\odot} \sim \theta_{12}$ have been determined with a remarkable
accuracy.

Oscillations of the atmospheric
$\nu_{\mu}$ ($\bar{\nu}_{\mu}$)
have been most clearly observed in the Super-Kamiokande
experiment. 
In particular the observed zenith angle dependence
of the multi-GeV and sub-GeV
$\mu$-like events \cite{SKatmo98,SKatmo03}
represents a compelling evidence.
(Indication for the atmospheric neutrino oscillations,
based mostly on the $\mu/e$ ratio, existed for a long time.) 
As is well known, the SK atmospheric
neutrino data is best described in
terms of dominant two-neutrino
$\nu_{\mu} \rightarrow \nu_{\tau}$
($\bar{\nu}_{\mu} \rightarrow \bar{\nu}_{\tau}$)
vacuum oscillations with maximal mixing.
The analysis thus fixes the parameters 
$|\Delta m^2_{31}| \sim |\Delta m^2_{32}| $ and 
$\theta_{atm} = \theta_{23}$ (since $\cos \theta_{13} \sim 1)$.

Finally, the remaining angle, $\theta_{13}$ remains
unknown, but is constrained from above by the reactor neutrino 
CHOOZ and Palo Verde experiments 
\cite{CHOOZ,PaloV}.
 
The Table \ref{tab:params} summarizes the present status of knowledge of the
oscillation parameters (assuming three mass eigenstates, i.e.
disregarding the possible existence of sterile neutrinos).

Thus, two of the three angles, and the two mass square differences
have been determined reasonably well. The unknown quantities, accessible
in future oscillation experiments (and discussed elsewhere in these
reports) are the angle $\theta_{13}$ and the sign of the 
$\Delta m_{32}^2 \sim \Delta m_{31}^2$.
If that sign is positive, the neutrino mass pattern is called a
{\it normal mass ordering} ($m_1 < m_2 < m_3$) and when it is negative
it is called {\it inverted mass ordering} ($m_3 < m_1 < m_2$). The extreme 
mass orderings, $m_1 < m_2 \ll m_3$ and $m_3 \ll m_1 < m_2$, are called the
{\it normal} and, respectively, {\it inverted} hierarchies.  
In addition, the phase $\delta$ governing
CP violation in the flavor oscillation experiments remains unknown,
and a topic of considerable interest. Determination of the CP phase
$\delta$ is again extensively discussed elsewhere in this report.

The remaining unknown quantities, 
the absolute neutrino mass scale, and the two
Majorana phases $\alpha_1$ and $\alpha_2$ are not accessible in
oscillation experiments. Their determination is the ultimate goal
of $0\nu\beta\beta$ and $\beta$ decay experiments.

\begin{table}[b]
\caption{Neutrino oscillation parameters determined from various experiments 
(2004 status)} 
\label{tab:params}
\begin{center}
\begin{tabular}{ccl}
Parameter & Value $\pm 1 \sigma$  & Comment\\
\hline 
$\Delta m_{21}^2$ & $8.2^{+0.6}_{-0.5} \times 10^{-5}$~eV$^2$  \\
 $ \theta_{12} $ & ${32.3^{\circ +2.7}_{-2.4}}$  & For $\theta_{13}=0$  \\
 $|\Delta m_{32}^2|$  & $2.0^{+0.6}_{-0.4} \times  10^{-3}$~eV$^2$  \\
  $\sin^2 2 \theta_{23} $  & $>0.94$ & For $\theta_{13}=0$\\
$\sin^2 2\theta_{13}$  & $<0.11$  & For $\Delta m^2_{atm} =
2\times 10^{-3}$ eV$^2$ \\
\end{tabular}
\end{center}
\end{table}

\subsection{Oscillations and direct neutrino mass measurements}

Direct neutrino mass measurements are based on 
the analysis of the kinematics of charged particles (leptons, pions) 
emitted together with
neutrinos (flavor states) in various weak decays.  The most sensitive
neutrino mass measurement to date, involving electron type neutrinos,
is based on fitting the shape of the beta spectrum (see section VI 
below). In such measurements the quantity
\begin{equation}
m_{\nu_e} = \sqrt{\sum_i |U_{ei}|^2 m_{\nu_i}^2}
\end{equation}
is determined or
constrained, where the sum is over all mass eigenvalues $m_{\nu_i}$
that are too close together to be resolved experimentally.

A limit on $m_{\nu_e}$ implies an {\it upper} limit on the {\it
minimum} value $m_{\nu_{min}}$ of all $m_{\nu_i}$, 
independent of the mixing
parameters $U_{ei}$: $m_{\nu_{min}} \leq m_{\nu_e}$,
i.e., the lightest neutrino cannot be heavier than
$m_{\nu_e}$. This is, in a sense, an almost
trivial statement.

However,  when the study of neutrino oscillations provides us with the values
of {\it all} neutrino mass-squared differences $\Delta m_{ij}^2$ 
(including their signs) and
the mixing parameters $|U_{ei}|^2$, 
and the value of $m^{2}_{\nu_e}$ has been determined
in a future experiment, 
then the individual neutrino mass squares
can be determined:
\begin{equation}
m_{\nu_j}^2 = m^{2}_{\nu_e} -  \sum_i |U_{ei}|^2
\Delta m_{ij}^2  ~~(\Delta m_{ij}^2 = m_{\nu_i}^2 - m_{\nu_j}^2) ~.
\end{equation}

On the other hand, if only the absolute values
$|\Delta m_{ij}^2|$
are known (but all of them), 
a limit on $m_{\nu_e}$ from beta decay may be used to
define an {\it upper} limit on the {\it maximum} value $m_{\nu_{max}}$
of $m_{\nu_i}$: 
\begin{equation}
m_{\nu_{max}}^2 \leq m^{2}_{\nu_e} + \sum_{i < j}
|\Delta m_{ij}^2|. 
\end{equation}
In other words, knowing  $|\Delta m_{ij}^2|$ one can use 
a limit on $m_{\nu_e}$ to constrain the heaviest
active neutrino.

\subsection{Oscillations, $0\nu\beta\beta$ decay, and neutrino mass}

The neutrinoless double beta decay,
\begin{equation}
(Z,A)  \rightarrow (Z+2,A) + e_1^- + e_2^-
\label{e:0nu}
\end{equation}
violates lepton number conservation. It can be recognized by
its electron sum energy spectrum. Since the nuclear masses 
are so much larger than the decay $Q$ value,
the nuclear recoil energy is negligible, and the electron sum energy
of the  $0\nu\beta\beta$ is simply a peak at $T_{e1} + T_{e2} = Q$
smeared only by the detector resolution.

The $0\nu\beta\beta$ decay involves a vertex changing two neutrons
into two protons with the emission of two electrons and nothing else.
One can visualize it by assuming that the process involves the exchange
of various virtual particles, e.g. light or heavy Majorana neutrinos,
right-handed weak interaction mediated by the $W_R$ boson, SUSY particles, etc.
No matter what the vertex is, the $0\nu\beta\beta$ decay
can proceed only when neutrinos are massive Majorana particles
\cite{SV82}. 

Of primary interest is the process mediated by the exchange
of light Majorana neutrinos interacting through the left-handed
$V - A$ weak currents. The decay rate is then,
\begin{equation}
[ T_{1/2}^{0\nu} (0^+ \rightarrow 0^+)]^{-1}
~=~ G^{0\nu}(E_0,Z)
\left| M_{GT}^{0\nu} - \frac{g_V^2}{g_A^2} M_F^{0\nu} \right|^2
\langle m_{\beta\beta} \rangle^2 ~,
\label{e:0nut}
\end{equation}
where $G^{0\nu}$ is the accurately calculable phase space integral,
$\langle m_{\beta\beta} \rangle$ is the effective neutrino mass,
and $M_{GT}^{0\nu}$,
$M_F^{0\nu}$ are the nuclear matrix elements.
The problems associated with the evaluation of
the nuclear matrix elements are discussed in Section IV,
where also corrections to the nuclear structure dependent part
of the decay rate are discussed.
 
If the $0\nu\beta\beta$ decay is observed, and the 
nuclear matrix elements are known, one can deduce the corresponding
$\langle m_{\beta\beta} \rangle$ value, which in turn is
related to the oscillation parameters by 
\begin{equation}
 \langle m_{\beta\beta} \rangle = 
| \sum_i |U_{ei}|^2 m_{\nu_i} e^{i\alpha_i} | ~,
\label{e:meff}
\end{equation}
where the sum is only over light neutrinos ($m_{\nu_i} < 10$ MeV). 
 The Majorana
phases $\alpha_i$ were defined earlier in Eq.(\ref{e:u3}). If the neutrinos
$\nu_i$ are $CP$ eigenstates, $\alpha_i$ is either 0 or $\pi$. 
Due to the presence of these unknown phases, 
cancellation of terms in the sum in
Eq.(\ref{e:meff}) is possible, 
and $\langle m_{\beta \beta} \rangle$ could be smaller than
any of the $m_{\nu_i}$ even if all neutrinos $\nu_i$ are
Majorana particles.

We can use the values in  
Table \ref{tab:params} and express the 
$\langle m_{\beta \beta} \rangle$ in terms of the three unknown
quantities: the mass scale, represented by the mass of the lightest
neutrino $m_{\nu_{min}}$, 
and the two Majorana phases. In doing so, it is useful
to distinguish three mass patterns: normal hierarchy (NH),
$m_1 < m_2 \ll m_3$ (i.e. $m_{\nu_{min}} = m_1$), inverted hierarchy (IH),
$m_3 \ll m_1 < m_2$  (i.e. $m_{\nu_{min}} = m_3$), and the 
quasi-degenerate spectrum (QD) where 
$m_{\nu_{min}} \gg \sqrt{|\Delta m_{32}^2|}$
as well as $m_{\nu_{min}} \gg \sqrt{|\Delta m_{21}^2|}$.

\begin{figure}
\begin{center}
\includegraphics[width=3. in]{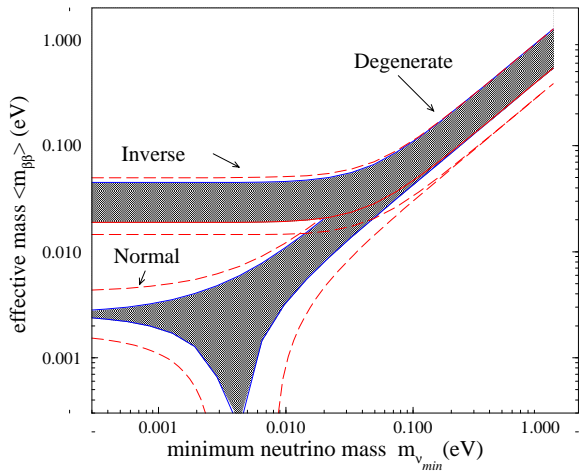}
\end{center}
\caption{Effective Majorana mass $\langle m_{\beta \beta} \rangle$
versus the minimum mass $m_{\nu_{min}}$. The different mass patterns
are indicated. The shaded region corresponds to the best values of 
oscillation parameters, and $\theta_{13} = 0$. The dashed lines 
indicate the expanded range corresponding to the 1$\sigma$ errors
of the oscillation parameters and the maximum allowed $\theta_{13}$.
Note that the uppermost line is unchaged (within this scale) in that
case.}
\label{fig:combp}
\end{figure}

In the case of normal hierarchy, and assuming that $m_{\nu_{min}} \equiv m_1$
can be neglected, one obtains 
\begin{equation}
\langle m_{\beta \beta} \rangle^{NH} \simeq
| \sqrt{\Delta m_{21}^2} \sin^2\theta_{12}\cos^2\theta_{13}
+  \sqrt{|\Delta m_{31}^2|} \sin^2\theta_{13} e^{-i\alpha_2} | ~.
\end{equation}
For $\theta_{13} = 0$ and the parameter values listed in 
Table \ref{tab:params},
$\langle m_{\beta \beta} \rangle^{NH} = 2.6 \pm 0.3$ meV.
On the other hand, if 
$\tan\theta_{13}^2 \ge \sin^2 \theta_{12} 
\sqrt{|\Delta m_{21}^2/ \Delta m_{31}^2} \sim 0.06$ a complete cancellation
might occur and $\langle m_{\beta \beta} \rangle^{NH}$ might be vanishingly
small. However, if $m_{\nu_{min}} > 0$ then
$\langle m_{\beta \beta} \rangle^{NH}$ may vanish even for
$\theta_{13} = 0$, see Fig. \ref{fig:combp}.

In the case of the inverted hierarchy, and again assuming that $m_{\nu_{min}} \equiv m_3$
can be neglected, one obtains 
\begin{equation}
\langle m_{\beta \beta} \rangle^{IH} \simeq
 \sqrt{|\Delta m_{31}^2|} \cos^2\theta_{13} 
\sqrt{1 - \sin^2 2\theta_{12} \sin^2 \frac{\alpha_2 - \alpha_1}{2}} ~.
\end{equation}
Thus, if $\theta_{13} = 0$ and for the parameter values listed in 
Table \ref{tab:params} 
$\langle m_{\beta \beta} \rangle^{NH} \simeq 14-51$ meV, depending on the
Majorana phases.

Finally, for the quasi-degenerate spectrum
\begin{equation}
\langle m_{\beta \beta} \rangle^{QD} \simeq
m_0 | (\cos^2 \theta_{12} e^{i \alpha_1} + \sin^2  \theta_{12} e^{i \alpha_2})
\cos^2 \theta_{13} + \sin^2 \theta_{13} | .
\end{equation}
 If $\theta_{13} = 0$ and for the parameters listed in 
Table \ref{tab:params}, $\langle m_{\beta \beta} \rangle^{QD} \simeq
(0.71 \pm 0.29)m_0$.

Detailed discussion of the relation between the 
$\langle m_{\beta \beta} \rangle$ and the absolute neutrino mass scale
can be found in numerous papers (see, e.g. some of the more recent
references \cite{BGGKP99,Viss99,Fer02,Bil01,Pas02,Pas04}).

In Fig.\ref{fig:combp} we show the plot of $\langle m_{\beta \beta} \rangle$ 
versus  $m_{\nu_{min}}$ using the oscillation parameters in 
Table \ref{tab:params}, and allowing for the maximum value of
$\theta_{13}$ and one $\sigma$ variations of them. One can clearly see
the three regions (NH, IH, QD). Thus, determination of the 
$\langle m_{\beta \beta} \rangle$ value would allow, in general,
to distinguish between these patterns, and to determine a range
of  $m_{\nu_{min}}$. One should keep in mind, however, that
there are caveats to this statement
for the situations where the corresponding
bands merge (e.g. the IH and QD  near $m_{\nu_{min}} \sim$ 0.05 eV).

Despite this caveats, obviously, if one can experimentally establish
that  $\langle m_{\beta \beta} \rangle \ge $ 50 meV, one can conclude
that the QD pattern is the correct one, and one can read off
an allowed range of $m_{\nu_{min}}$ values
from the figure. (The sign of $\Delta m_{31}^2 \sim \Delta m_{32}^2$
will remain undetermined in that case, however.)

On the other hand, if  
$\langle m_{\beta \beta} \rangle \sim$ 20-50 meV only an upper limit
for the $m_{\nu_{min}}$ can be established, and the pattern is likely
IH, even though exceptions exist. However, if (and that is unlikely
in a foreseable future) the value of $m_{\nu_{min}}$ can be determined
independently, the pattern can be resolved.

Finally, if one could determine that
$\langle m_{\beta \beta} \rangle \le$ 10 meV but nonvanishing
(which is again is unlikely in a foreseable future), one could conclude
that the NH pattern is the correct one.

Altogether, observation of the \BBz\ decay, and accurate determination of
the $\langle m_{\beta \beta} \rangle$ value would not only establish
that neutrinos are massive Majorana particles, but would contribute
considerably to the determination of the absolute neutrino mass scale. 
Moreover, if the neutrino mass scale would be known from independent
measurements, one could possibly obtain from the measured 
$\langle m_{\beta \beta} \rangle$ also some information about the
CP violating Majorana phases. 

\subsection{Absolute neutrino mass scale}

As shown above, \BBz\ and $\beta$ decays both depend on different
combinations of the neutrino mass values and oscillation parameters.
The  \BBz\ decay rate is proportional to the square of
a coherent sum of the Majorana neutrino masses because
\mee\
arises from exchange of a virtual neutrino. On the other hand,
in beta decay one can determine an
incoherent sum because a real neutrino is emitted.

Quite different source of information is based on cosmological
and astrophysical observations where the density of 
the primordial neutrino sea is determined or constrained and thus a
parameter proportional to the sum of the neutrino masses
is determined.

Massive neutrinos would contribute to the cosmological
matter density an amount,
\begin{equation}
\Omega_{\nu} h^2  = \Sigma m_{\nu_i}/ 92.5 {\rm ~eV} ~,
\end{equation}
where $\Omega_{\nu}$ is the neutrino mass density relative to the critical
density and $100 h$ is the Hubble constant in km/s/Mpc. 
From the requirement that the neutrinos left over from the
Big Bang do not overclose the universe  an upper limit,
with a minimum assumptions (essentially just the requirement of
stability), is obtained
\begin{equation}
m_{\nu} \le \frac{46  {\rm ~eV}}{N_{\nu}} ~,
\end{equation}
where $N_{\nu}$ is the number of neutrino species with 
standard weak interactions \cite{Hann04}.

More restrictive limits are obtained from the requirement that
excessive free streaming in the early universe would not
suppress small scale power of the observed matter distribution.
The relation between the damping scale $d_{FS}$ caused by
free streaming, and the neutrino mass is approximately
\begin{equation}
d_{FS} {\rm ~(Gpc)} \sim 1/m_{\nu} {\rm ~(eV)} ~.
\end{equation}
The data on Cosmic Microwave Background (CMB) and large scale
galaxy surveys can be used to constrain $N_{\nu}m_{\nu}$ for the 
quasi-degenerate neutrino mass spectrum, and thus also $m_{\nu}$
for various assumed number of neutrino flavors
$N_{\nu}$. The following Table \ref{tab:cosm}
is based on \cite{Raff03}. Different analyses with different assumptions
typically reach similar conclusions, suggesting that these limits
are fairly robust (see more discussion further in this report).

\begin{table}
\caption{Limits on  $N_{\nu}m_{\nu}$ and on $m_{\nu}$
assuming quasi-degenerate mass spectrum. In the last column
is the heaviest neutrino mass assuming that one neutrino
dominates the sum. } 
\begin{center}
\label{tab:cosm}
\begin{tabular}{|cccc|}
 $N_{\nu}$ & $N_{\nu}m_{\nu}$ (eV)  & $m_{\nu}$ (eV) &  $m_{\nu_{max}}$ (eV) \\
\hline 
3 & 1.01 & 0.34 & 0.73 \\
4 & 1.38 & 0.35  & 1.05 \\
5 & 2.12 & 0.42 & 2.47 \\
6 & 2.69 & 0.45 & 4.13 \\
\end{tabular}
\end{center}
\end{table}

For completeness, note that, in principle, 
neutrino mass can be also extracted from
the time of flight determination of neutrinos from future galactic
supernova. However, one does not expect to be able to reach
sub-eV sensitivity with this method (see e.g.\cite{Beacom98}). 

\begin{figure}
\vspace{9pt}
\begin{center}
\includegraphics[width=3.0 in]{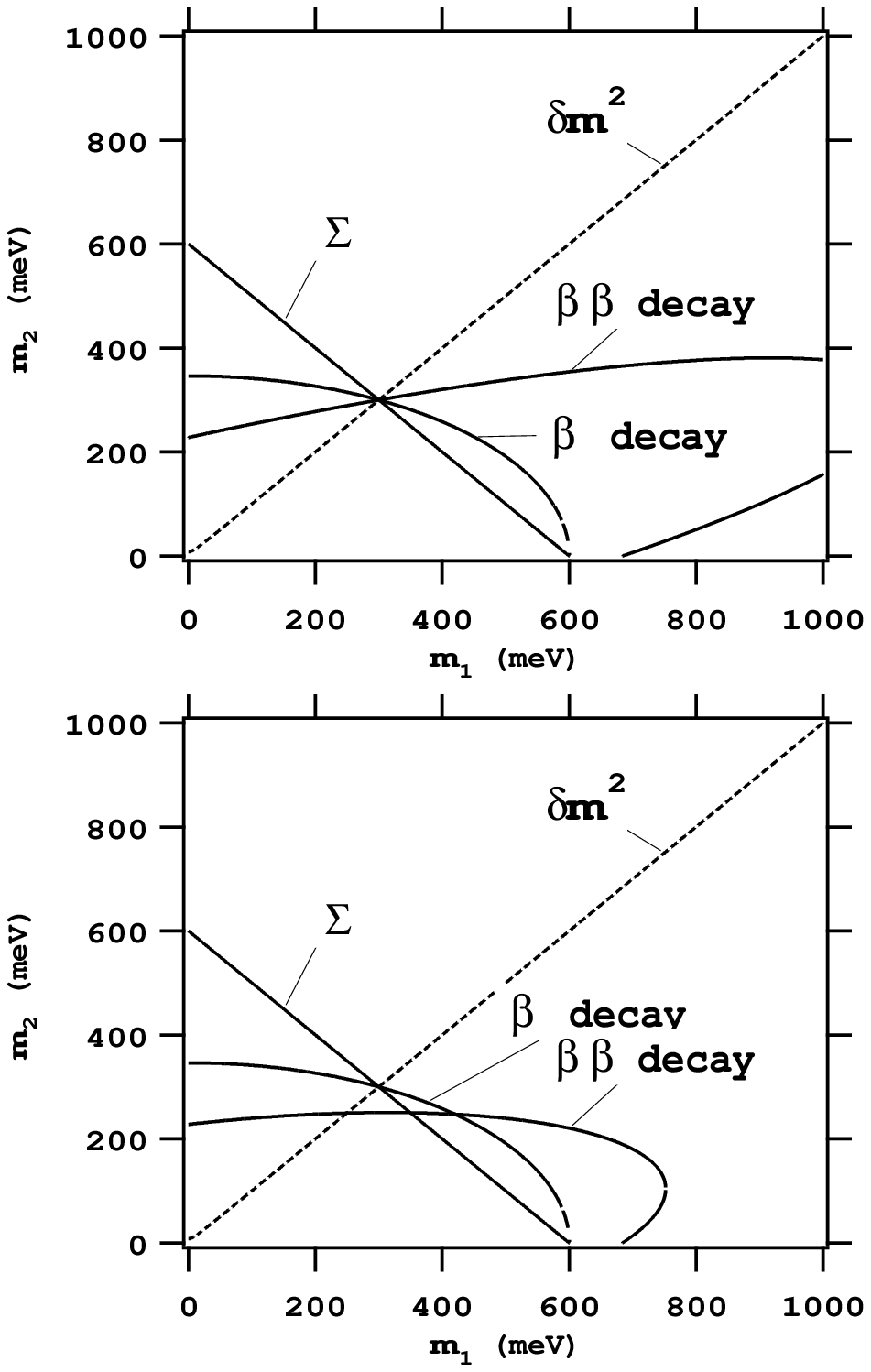}
\end{center}
\caption{A consistency plot for the neutrino mass eigenvalues $m_1$
and $m_2$, for various
hypothetical measurements. This set of curves indicates how measured
values of $\Sigma$,
\mee, \ms, and \mb\ constrain the mass eigenvalues. The \BBz\ curve
has been drawn for an {\em incorrect}
value of the phase in the bottom panel to indicate the sensitivity of
this technique for extracting the CP-violating phase.
See text for further description of the parameters
used to draw the curves.}
\label{fig:masses}
\end{figure}

It is worthwhile to stress that the various methods that depend 
on the neutrino absolute mass scale are complementary. If, ideally,
a positive measurement is reached in all of them (\BBz\, $\beta$ decay,
cosmology) one can test the results for consistency and perhaps determine
the Majorana phases. We illustrate the idea \cite{ELL04} in 
Fig. \ref{fig:masses}
using a two-neutrino-species example of such a set of measurements. 
(A 3-species example is discussed in Ref. \cite{ELL04}.) We took the mixing
matrix and $\Delta m^2$ to be the best fit to the
solar-neutrino data, with an arbitrary
value for the Majorana phase $\alpha$
(of which there is only one) of 2.5 radians.  We then made up values
for $\Sigma$, \mee, and \mb\, assuming them to be the results of pretend
measurements.  Each curve in
the $m_2$ vs.\ $m_1$ graph is defined by
one of these measurements.  We chose the value of $\Sigma$ (from cosmology)
to be 600 meV, corresponding to a quasidegenerate hierarchy, and let
\mb\ = 300 meV and \mee\ = 171 meV.
The $m_2$ versus $m_1$ curves from the ``measurements" of the
oscillation parameters, $\Sigma$, and $\beta$ decay are:
\begin{eqnarray}
  m_2 & = & \Sigma - m_1 \nonumber \\
  m_2 & = & \sqrt{m_1^2 + \delta m_{12}^{2}}  \nonumber \\
  m_2 & = & \sqrt{ (\langle m_{\beta} \rangle/U_{e2})^2 - (m_1
U_{e1}/U_{e2})^2}~.
\end{eqnarray}

The \BB\ constraint is a little more complicated than that from
$\beta$ decay. The curve is also
an ellipse but rotated with respect to the axes.
All of the equations express $m_2$ in terms of $m_1$ and measured
parameters and all should
intersect at one ($m_1$,$m_2$) point.
 However, because the point is
overdetermined, the \BB\ ellipse will
intersect only for a correct choice of $\alpha$.  This provides a way to
determine $\alpha$.  In Figure \ref{fig:masses} we drew the \BB\
ellipse for $\alpha = 2.0$
radians and for the ``true" value of 2.5 radians
to show how the intersection does indeed depend on a correct choice
of the phase.

Many authors have examined the potential to combine measurements from
\BB\ decay, tritium $\beta$
decay, and cosmology to determine the Majorana phases. (See e.g.
\cite{ROD01,Pasc03,Abada03}.)

\subsection{Leptogenesis}

Leptogenesis is a natural possibility when the see-saw mechanism is 
realized by Nature. Since this mechanism predicts the light neutrinos to 
be of Majoarana type,  \BBz\ is also expected. 
Though the details of the connection of the high and low energy parameters 
remain to be investigated in a given model, leptogenesis always comes with 
\BBz. 
The only exception is when in the basis in which the charged leptons are 
diagonal the neutrino mass matrix has a zero entry in the ee element, 
resulting in no neutrinoless double beta decay.  
This implied texture will however have interesting physical implications 
by itself. In this section we discuss the relationship between neutrino mass and
leptogenesis.

The origin of the matter-antimatter asymmetry is one 
of the most important questions in cosmology.
The  presently observed baryon asymmetry is~\cite{wmap}
\begin{equation}
Y_B=\frac{n_B-n_{\bar B}}{s}\simeq 6.5 \times 10^{-10}\,.
\end{equation}
In 1967 A. Sakharov suggested that the baryon density
can be explained in terms of microphysical laws~\cite{sakharov}.
Three conditions need to be fullfilled:
\begin{itemize}
\item 
Baryon number (or Lepton number, for the leptogenesis mechanism) violation.
Assuming that the initial baryon asymmetry is negligible, as implied by inflationary models, 
baryon number violation is required 
to produce a net final baryon asymmetry.
\item
C and CP violation. The CP symmetry is the product of charge conjugation 
and parity. If CP is conserved, every reaction which produces a 
particle will be accompanied by a reaction which gives its antiparticle,
with no creation of a net baryon number.
\item
Departure from thermal equilibrium.
\end{itemize}

Several mechanism have been proposed to understand the baryon asymmetry.

 - Planck-scale baryogenesis. 
While we can expect a quantum theory of gravity 
not to preserve any quantum number, these effects would be relevant 
only at very early times and the baryon asymmetry generated
in this way would be subsequently diluted in the inflationary epoch.

- GUT baryogenesis (see, e.g., ref.~\cite{kolb90}).
GUT theories assume the existence of an underlying theory 
at high scale with a simple gauge group which is then broken 
to the Standard Model gauge group at low energy scale.
The unification scale is expected to be at $\sim 10^{16}$ GeV. 
In general these models have lepton and baryon number violation and
have many new parameters with respect to the Standard Model
which can provide new sources of CP-violation.
However cosmological considerations, mainly 
overproduction of gravitinos, suggest that 
the reheating temperature after inflation 
should be not higher than $\sim 10^9$ GeV. 
If so, any baryon asymmetry produced at the GUT scale
would be later diluted and negligible.

- Electroweak baryogenesis.
The interactions of the Standard Model, even if preserving the lepton and baryon number
at the perturbative level, do violate  it non-perturbatively~\cite{thooft}.
The ground states of a non-abelian gauge group
are labeled by an integer $n$ and are separated 
by energy barriers one from the other. 
At zero temperature,
the instantons transitions between different vacua
are exponentially suppressed,
because they are due to quantum-tunneling effects. 
At high temperatures above the electroweak breaking scale,
these transitions
can proceed through sphaleron processes.
This produces a  net baryon number.
It was pointed out the Standard Model
can satisfy all the three Sakharov condition for baryogenesis~\cite{kuzmin85}.
The departure from equilibrium is realized at the electroweak
symmetry breaking scale when 
the Higgs field acquires a vacuum expectation value different from zero
and a phase transition takes place.
However, the generated baryon asymmetry 
is by far too small to account for the observations.
This is mainly due to the fact that the departure from equilibrium
is not strong enough,
as implied by a 
heavy Higgs, $m_h > 80$ GeV.\\
Supersymmetric extensions of the Standard Model
contain new sources of CP-violation
and an enlarged set of parameters which might produce
a first order phase transition.
The parameter space which allows
for a sizable baryon number generation
is rather small (it requires in particular a light stop)
and it will be soon tested.

- Affleck-Dine baryogenesis~\cite{affleck85}.
Scalar fields carrying baryon and lepton number are naturally present 
in supersymmetric extension of the Standard Model
(e.g. squarks and sleptons).
A coherent scalar field, a field which has large vacuum expectation value,
can in principle carry a large baryon number.
This condensate can be produced in the Early Universe by the evolution
of the scalar field along flat directions of its potential.
The decay of the condensate into ordinary matter is then responsible 
for the baryon asymmetry of the Universe,
as the baryon charge of the condensate is transferred 
to quarks. 
This is still a viable explanation for the baryon asymmetry of the Universe.

- Leptogenesis~\cite{lepto}. 
Let us notice that $B-L$ is conserved
both at the perturbative and non-perturbative level.
This implies that if one creates a net $B-L$,
(e.g., a lepton number),
the sphaleron processes would leave 
both baryon and lepton number comparable 
to the original $B-L$. 
This idea is implemented in the leptogenesis scenario~\cite{lepto}.
Leptogenesis is particularly appealing because it takes place
in the context of see-saw models~\cite{seesaw}, 
which naturally  explain the smallness
of neutrino masses.
The see-saw mechanism
requires the existence of heavy right-handed (RH) Majorana neutrinos, 
completely neutral 
under the Standard Theory gauge symmetry group.
Consequently, they can acquire Majorana masses that are not related to the electroweak
symmetry breaking mechanism, and can, in principle, be much heavier than any of the known
particles.
Introducing a Dirac neutrino mass 
term and a Majorana mass
term for the right--handed   
neutrinos via the Lagrangian 
\begin{equation} \label{eq:L}
 -{\cal{L}} = \overline{\nu_{Li}} \, (m_D)_{ij} \, N_{Rj} \, + 
\frac{1}{2} \, \overline{(N_{Ri})^c} \, (M_R)_{ij} \, N_{Rj} ~, 
\end{equation}
%
leads, for sufficiently large $M_R$, 
to the well know see--saw \cite{seesaw} formula 
\begin{eqnarray} \label{eq:seesaw}
m_\nu  &   \simeq - m_D \ M_R^{-1} \ m_D^T~, \\
  & = U_{\rm PMNS} \   D_m  \ U_{\rm PMNS}^T
\end{eqnarray}
%
where terms of order  ${\cal O}(M_R^{-2})$ 
are neglected.
Here $D_m$ is a diagonal 
matrix containing the masses $m_{1,2,3}$
of the three light massive Majorana neutrinos and 
$U_{\rm PMNS}$ is the unitary 
Pontecorvo--Maki--Nagakawa--Sakata 
lepton mixing matrix.

    The CP-violating and out-of-equilibrium
decays of RH neutrinos produce a lepton asymmetry \cite{lepto}
that can be  converted into a baryon asymmetry
through anomalous electroweak processes~\cite{kuzmin85,conv}. 
The requisite $CP$ violating decay asymmetry 
is caused by the interference of 
the tree level contribution and 
the one--loop corrections in the decay rate
of  the three heavy 
Majorana neutrinos, 
$N_i \rightarrow \Phi^- \, \ell^+$ and $N_i \rightarrow \Phi^+ \, \ell^-$:
\begin{equation} \label{eq:eps}
\begin{array}{ll} 
\varepsilon_i & = \frac{\displaystyle \Gamma (N_i \rightarrow \Phi^- \, \ell^+) - 
\Gamma (N_i \rightarrow \Phi^+ \, \ell^-)}{\displaystyle \Gamma (N_i \rightarrow \Phi^- \, \ell^+) +  
\Gamma (N_i \rightarrow \Phi^+ \, \ell^-)} \\
&  \simeq  \displaystyle \frac{\displaystyle 1}{\displaystyle 8 \, \pi \, v^2} 
\sum_{j\neq i} \frac{{\rm Im} (m_D^\dagger m_D)^2_{ij}}{(m_D^\dagger m_D)_{ii}} \, \left( f(x_j) + 
g(x_j) \right)~,
\end{array} 
\end{equation}
%
where $\Phi$ and $\ell$ indicate the Higgs field and 
the charged leptons, respectively.
Here $v \simeq 174$ GeV is the electroweak symmetry breaking 
scale and $x_j \equiv M_j^2 / M_i^2$. The functions $f$ and $g$ stem from vertex 
\cite{lepto,vertex} and  from self--energy \cite{self} contributions. 
\begin{equation} \label{eq:fapprox}
\begin{array}{ll} 
f(x) = \sqrt{x} \left(1 - 
(1 + x) \, \ln \left(\frac{\displaystyle 1 + x}{\displaystyle x}\right) \right) \\
g(x) = \frac{\displaystyle \sqrt{x}}{\displaystyle 1 - x}
\end{array} \, . 
\end{equation}
%
For $x \gg 1$, i.e., for hierarchical 
heavy Majorana neutrinos, 
one has $f(x) + g(x) \simeq -\frac{3}{2\sqrt{x}}$. 
Then, the baryon asymmetry is obtained via 
$Y_B = a \, (\kappa /g^\ast) \, \varepsilon_1$,
where $a \simeq -1/2$ is the fraction of 
the lepton asymmetry converted into 
a baryon asymmetry \cite{kuzmin85,conv}, 
$g^\ast \simeq 100$ is the number of massless 
degrees of freedom at the time of the decay, 
and $\kappa$ is an efficiency factor that is 
obtained by solving the Boltzmann equations~\cite{buch}. 
Typically, one gets $Y_B \sim 6 \times 10^{-10}$ when 
$\varepsilon_1 \sim (10^{-6} - 10^{-7})$ and 
$\kappa \sim (10^{-3} - 10^{-2})$. 
A similar estimate of $Y_B$ 
is obtained in the supersymmetric (SUSY) 
theories as well~\cite{self}.

In non-supersyymetric models which embedd 
the see-saw mechanism,
lepton flavour violating (LFV)
charged lepton decays such as
$\mu \rightarrow e + \gamma$, $\tau \rightarrow \mu + \gamma$,
$\tau \rightarrow e + \gamma$,
are predicted to take place with branching ratios
which are strongly suppressed~\cite{petcov77}.
SUSY theories have additional sources of lepton charge non-conservation. 
If SUSY is broken above the RH
Majorana mass scale, as, e.g, in gravity-mediated breaking scenarios, 
there are renormalization group effects 
that generate new lepton charge non-conserving couplings at low energy
even if such couplings are absent at the GUT scale~\cite{borz,iba}. 
In contrast to the non-supersymmetric
case, the LVF processes can
proceed with rates and cross-sections
 which are within the sensitivity of presently operating
and proposed experiments.
Under the assumption of flavour universality 
of the SUSY breaking sector (scalar masses 
and trilinear couplings) at the GUT scale $M_X$, 
using 
the leading--log approximation,
one can estimate ~\cite{borz,iba,petmu} 
the branching ratio of the 
charged lepton decays 
$\ell_i \rightarrow \ell_j + \gamma$,
$\ell_i(\ell_j) = \tau, \mu, e$ for $i(j) = 3,2,1$,
$i > j$,
\begin{equation} \label{eq:muegBR}
\begin{array}{ll}
BR(\ell_i \rightarrow \ell_j + \gamma) \simeq 
\alpha^3 \, \left( \frac{\displaystyle (3 + a_0) \, m_0^2}
{\displaystyle 8 \, \pi^2 \, m_S^4 \, G_F \, v^2} \right)^2  \\
\left|(m_D  \ 
{\rm diag}(\log \frac{M_X}{M_{1}},\frac{M_X}{M_{2}},\frac{M_X}{M_{3}}) 
\  m_D^\dagger)_{ij}\right|^2 \tan^2 \beta~,
\end{array}
\end{equation}
%
where $m_S$ denotes a slepton mass, $m_0$ is the universal 
mass scale of the sparticles and $a_0$ is a trilinear 
coupling (all at $M_X$) 
(for details, see, e.g., ref.~\cite{petmu}). 

Establishing a connection between the parameters
at low energy (neutrino masses, mixing angles and CP-violating phases),
measurable in principle in present and future experiments,
and at high energy (relevant in leptogenesis) has gathered
a great interest in the last few years.
The number of parameters in the full Lagrangian
of models which implement the see-saw mechanism
is larger than the ones in the low-energy sector: 
in the case of 3 light neutrinos and three heavy ones,
at high energy the theory contains in the neutrino sector 18 parameters
of which 12 real ones and 6 phases, while at low energy
only 9 are accessible - 3 angles, 3 masses and 3 phases.
The decoupling of the heavy right-handed neutrinos 
implies the loss of information on 9 parameters.
This implies that reconstructing the high energy
parameters entering in the see-saw models from the measurement
of the masses, angles and CP-violating phases of $m_\nu$
is in general difficult, if not impossible,
and depends on the specific model considered.

Using the weak basis in which both
$M_R$ and the charged lepton mass matrix are real and diagonal,
we find useful to parametrize the Dirac mass by the
biunitary or the orthogonal parametrizations.

{\bf Biunitary parametrization.}
We can write the complex $3 \times 3$ Dirac mass as (see, e.g., ref~\cite{PPRlepto}):
\begin{equation} \label{eq:mdULUR}
m_D = U_L^\dagger \, m_D^{\rm diag} \, U_R~,  
\end{equation}
%
where $U_L$ and $U_R$ are 
unitary $3 \times 3$ matrices and  
$m_D^{\rm diag}$ is a real diagonal matrix.
All the CP-violating phases are contained in 
$U_L$ and $U_R$.

{\bf Orthogonal parametrization.}

By using the see-saw formula, eq.~(\ref{eq:seesaw}),
we can express $m_D$ as~\cite{iba,PPYaguna}:
\begin{equation}
 \label{eq:mdO}
m_D =  i \, U_{\rm PMNS} \, D_m^{1/2}  \, R \, M_R^{1/2}~,
\end{equation}
where $D_m$ is the diagonal real matrix 
which containes the low-energy light neutrino masses,
and  $R$ is a complex orthogonal matrix. 
$R$ contains 3 real parameters 
and 3 phases.

The use of the two indicated parametrization clarifies 
the dependence of  the processes
we are interested in, e.g. leptogenesis and 
LFV charged lepton decays, on the different parameters 
entering in $m_D$.
In particular we have that:\\
- for leptogenesis, 
the decay asymmetry $\varepsilon_1$ 
depends  on the hermitian matrix 
$m_D^\dagger m_D$:
\begin{equation} \label{eq:mddmd}
m_D^\dagger m_D = 
\left\{ 
\begin{array}{cc} 
U_R^\dagger \, (m_D^{\rm diag})^2 \, U_R~, & \mbox{bi--unitary;} \\[0.3cm]
 M_R^{1/2} \, R^\dagger \, D_m \, R \, M_R^{1/2}~,
& {\rm orthogonal}. \\
\end{array}
\right. 
\end{equation}
%
We can notice that the PMNS unitary mixing matrix $U$
does not enter esplicitly into the 
expression for the lepton asymmetry.

- Concerning LFV decays,
in supersymmetric versions of the see--saw mechanism, 
their rates 
depend approximately
on \cite{borz,iba} 
\begin{equation} \label{eq:mdmdd}
m_D m_D^\dagger = 
\left\{ 
\begin{array}{cc} 
U_L^\dagger \, (m_D^{\rm diag})^2 \, U_L~, \\[0.3cm]
 U_{\rm PMNS} \, 
D_m^{1/2}  \, R \, M_R \, R^\dagger \,D_m^{1/2} 
\, U_{\rm PMNS}^\dagger~, \\
\end{array}  \right .
\end{equation}
by using the biunitary (upper expression)
or the orthogonal parametrization (lower expression).

- finally, in the biunitary parametrization,
the neutrino mass matrix $m_\nu$
can be written as:
\begin{equation} \label{eq:mnupara}
m_\nu = 
- U_L^\dagger \, m_D^{\rm diag} \, U_R \, M_R^{-1} \,  U_R^T \, 
m_D^{\rm diag} \, U_L^\ast~.
\end{equation}

This shows that 
the phases in $U_{\rm PMNS}$
receive contribution from CP-violation both in the right-handed sector,
responsible for leptogenesis, and in the left-handed one, which enters
in lepton flavour violating processes. 
Due to the complicated way in which the high energy phases and real parameters
enter in $m_\nu$, eq.~(\ref{eq:mnupara}), 
if there is CP-violation at high energy, as required by the leptogenesis mechanism,
we can expect 
in general to have CP-violation at low-energy,
as a complete cancellation would require some fine-tuning 
or special forms of $m_D$ and $M_R$.
Let us mention that an observation of CP-violation at low energy, however,
does not imply necessarily CP-violation in $U_R$ 
as it might receive all its contributions
from $U_L$. 

More specifically, from eq.~(\ref{eq:mnupara}), we see that,
 in general, there is no one-to-one link
 between low energy CP-violation
in the lepton sector and the baryon asymmetry: 
a measurement of the low energy CP-violating phases
does not allow to reconstruct the leptogenesis phases~\cite{BRO01}.
However most specific models
allow for such a connection. 
In particular if the number of parameter is reduced 
in $m_D$, then a one-to-one correspondence 
between high energy and low energy parameters
might be established. 
This can be achieved in models which allow for CP-violation 
only in the right-handed sector, that is in $U_R$,
or which reduce the number of independent parameters
at high energy, for example 
by requiring only two right-handed neutrinos~\cite{lepto2nu}.
Each model of neutrino mass generation
should be studied in detail separately to establish 
the feasibility of the leptogenesis mechanism~\cite{leptoconnection}. 

Furthermore the amount of baryon asymmetry depends  
on the type of light and heavy neutrino mass spectra
and succesful leptogenesis imposes some constrains
on the allowed mass scales~\cite{leptoheavy,dibari,PPRlepto}.
 For example in ref.~\cite{PPRlepto} 
the case of strongly hierarchical see-saw models was analyzed,
assuming a hierarchical structure of
the heavy Majorana neutrino
masses and of the neutrino Dirac mass matrix
$m_D$. 
In order to produce a sufficient amount of 
baryon asymmetry via the leptogenesis  
mechanism, the scale of $m_D$ should be given by the  
up--quark masses. In this case the $CP$ violation effects in neutrino 
oscillations can be observable,
but, in general, there is no direct connection 
between the latter and the $CP$ violation
in leptogenesis.
In ref.~\cite{dibari} it is also shown that, in the case of thermal leptogenesis
with a hierarchical structure of the heavy Majorana masses,
a strong bound on the light neutrino masses can be put
of order $0.1$~eV. 
This bound can be weakened if one assumes non-thermal
right-handed neutrino production
and/or an enhancement of the decay asymmetry
due to quasi-degenerate right-handed neutrinos 
(resonant leptogenesis).  
For example, in ref.~\cite{PPYaguna},
the case of quasi-degenerate spectrum 
of light neutrinos, 
$m_{1,2,3} \cong m_{\nu} > 0.2$~eV,
was considered.
It was shown that leptogenesis is feasible and can provide
the observationally required baryon asymmetry of the Universe.
Furthermore it was pointed out that the branching ratios for the lepton flavour processes,
$\mu \rightarrow e + \gamma$, 
$\tau \rightarrow \mu + \gamma$ and 
$\tau \rightarrow e + \gamma$ decay,
might be close to the present bounds
and that they depend on the Majorana CP-violating phases
which enter in \mbox{$\langle m \rangle$}.

\vspace{2mm}

The possible observation of \BBz\ decay
would play an important role in understanding the origin
of the baryon asymmetry
as it would imply that lepton number 
(one of the main conditions for leptogenesis)
indeed is not conserved. 
Furthermore the Majorana nature of neutrinos would be established:
the see-saw mechanism would be regarded as a reasonable
explanation of neutrino mass generation. Leptogenesis naturally
takes place in this scenario.
Finally the observation of CP-violation in the lepton sector,
in neutrino oscillation experiments and/or \BBz\ decay,
would suggest the existence of CP-violation at high energy,
which might be related to the one responsible 
for leptogenesis.

In conclusion, the observation of lepton number violation in \BBz\ decay
and, in addition, possibly of CP-violation in the lepton sector,
would be a strong indication, even if not a proof (as 
it is not possible to reconstruct in a model indipendent way 
the high energy parameters from $m_\nu$),
of leptogenesis
as the explanation for the observed baryon asymmetry of the Universe.

\subsection{Alternative tests of lepton number conservation}

Our main emphasis is on the study of the \BBz\ decay,
\begin{equation}
(A,Z)   \longrightarrow  (A,Z+2)\; +\; e^-\; +\; e^- ~. 
\end{equation}
However, alternatively, similar physics could be tested in other
processes, e.g. in
\begin{eqnarray}
\mu^- \; +\; (A,Z) & \longrightarrow & e^+ \;+\; (A,Z-2) \label{mu_e}\\
K ^+       & \longrightarrow & \mu^+ \; +\;  \mu^+ \; +\; \pi^- \label{k_plus}
\end{eqnarray}
which can be mediated by the exchange of virtual massive Majorana
neutrinos~\cite{ROD00,zuber00} and where rather restrictive limits
on the corresponding branching ratios exist. 
These Lepton number violating decay
rates are typically proportional to an average Majorana
neutrino mass of the form:
$\langle m_{\ell \ell '}\rangle^2= 
|\sum_i U_{\ell i} U_{\ell ' i} e^{\alpha_i} m_i|^2$,
where $\alpha_i/2$ denotes the Majorana phases (which can lead to destructive
interference).

Searches for muon-positron conversion~(\ref{mu_e}) and rare kaon decays
~(\ref{k_plus}) yield: $\langle m_{\mu e}\rangle < 17(82)$ MeV,
and $\langle m_{\mu \mu}\rangle < 4\times 10^4$ MeV,
respectively (see ~\cite{ELL02} and references therein).
These bounds therefore do  not constitute a useful constraint for the mass
pattern of the three light neutrinos. Similarly to the kinematic tests,
new Majorana mass experiments are focusing on process \BBz\,
offering  by far the best sensitivity.

Note that other lepton number violating processes might also exist,
causing for example the emission of the $\bar{\nu}_e$ antineutrinos
by the Sun. Even though the existence of such processes would again
signal that neutrinos are massive Majorana particles, the interpretation
of the rate in terms of neutrino masses is not straigthforward. Instead,
other mechanisms might be at play, like the spin-flavor precession 
combined with neutrino oscillations (relevant parameter is then
$\mu_{\nu} B_T$, where $\mu_{\nu}$ is the transition magnetic
moment and $B_T$ is the magnetic field in the Sun 
\cite{Akh88}) or neutrino
decay (the relevant parameter is then $\tau/m$ of the heavier neutrino
\cite{He88}).

\subsection{Mechanism of \BBz\ decay}

Although the occurrence of \BBz\ decay implies the existence of
massive Majorana neutrinos \cite{SV82}, 
their exchange need not be the only and even dominant
contribution to the decay rate.  Almost any physics that violates the total lepton
number can cause \BBz\ decay.    A heavy Majorana neutrino can be
exchanged, or supersymmetric particles, or a leptoquark. Right-handed
weak currents, either leptonic or hadronic, can cause the absorption
of an emitted virtual neutrino without the helicity flip that depends
on the neutrino mass. 
Moreover, heavy particle exchange, heavy neutrinos,
supersymmetric particles (see, e.g., \cite{heavy}),
resulting from lepton number violation dynamics at some scale above the
electroweak one can lead to \BBz\ decay with the same single
electron spectra as in the favored case of the light Majorana 
neutrino exchange.
These alternative mechanisms, if dominant,
 would not allow (or would make it very difficult) to extract the
effective Majorana mass $\langle m_{\beta\beta} \rangle$ from the measured \BBz\ decay rate.

The relative size of heavy ($A_H$) versus light
particle ($A_L$) exchange contributions to the decay amplitude
can be crudely estimated as follows~\cite{Moh98}:
\begin{equation}
A_L \sim G_F^2  \frac{m_{\beta \beta}}{\langle \bar{k}^2 \rangle}  ,~  
 A_H \sim G_F^2  \frac{M_W^4}{\Lambda^5}  ,~
\frac{A_H}{A_L} \sim \frac{M_W^4 \langle \bar{k}^2 \rangle } 
{\Lambda^5  m_{\beta \beta} }  \ ,
\end{equation}
where $\bar{k} \sim ( 50 \ {\rm MeV} ) $ is the
typical light neutrino virtual momentum, and $\Lambda$ is the heavy mass
scale relevant to the lepton number violation dynamics.
Therefore,  $A_H/A_L \sim O(1)$ for  $m_{\beta \beta} \sim 0.1-0.5$
eV and $\Lambda \sim 1$ TeV, and  thus the TeV
scale leads to similar $0 \nu \beta \beta$ decay rate as the
exchange of light Majorana neutrinos with the effective mass
$m_{\beta \beta} \sim 0.1-0.5$ eV. (Note that if this heavy
mass scale $\Lambda$ is larger than 3 - 5 TeV the $A_L$
will clearly dominate.)

The exchange of heavy particles involves short-range propagators
that give rise to decay rates
of the same form as in the mass-driven mode:  simple
two-s-wave-electron phase space multiplied by the square of an
amplitude. Two-nucleon correlations do not suppress the effects of
heavy particles, which can be transmitted between nucleons by pions
\cite{heavy}.
The angular distributions and single-electrons spectra will
therefore be the same in all these processes.
One way to
distinguish one from another is to take advantage of the different
nuclear matrix elements that enter the amplitudes (leading to
different total decay rates).  Unknown parameters such  as the
effective neutrino mass or the trilinear $R$-parity-violating
supersymmetric coupling (violation of $R$ parity naturally accompanies
Majorana neutrino-mass terms) also enter the rates, so several transitions
would have to be measured.
This might be best accomplished  \cite{SF02}  by measuring
transitions to several final states in the same nucleus, but if the
matrix elements can be calculated accurately enough one could also
measure the rates to the ground states of several different nuclei.
The problems in determining the source of \BBz\ decay are mitigated by
constraints from other
experiments on many extra-Standard models.  Some of these constraints
will be much stronger once
the Large Hadron Collider comes on line.

Alternatively, it has been argued in \cite{Ci04} that the study
of the lepton flavor violating processes $\mu \rightarrow e$ conversion,
and $\mu \rightarrow e + \gamma$ will provide important insight
about the mechanism of the \BBz\ decay. In particular, the ratio
of the branching fractions of these processes provides a diagnostic
tool for establishing whether the exchange of light Majorana neutrinos
is the dominant mechanism (and hence determination of \mee\ is possible)
or whether further more involved analysis is needed.

\section{III. Nuclear Structure Issues}

The observation of \BBz\ decay would immediately tell us that neutrinos are
Majorana particles and give us an estimate of their overall mass 
scale. But without
accurate calculations of the nuclear matrix
elements that determine the decay rate it will be difficult to reach 
quantitative
conclusions about masses and hierarchies.  

Theorists have tried hard to develop
many-body techniques that will allow such calculations.  
In order to test the calculations they have tried to
calibrate them against related observables:  \BBt\ decay, ordinary
$\beta^+$ and $\beta^-$ decay, Gamow-Teller strength distributions, odd-even
mass differences and single-particle spectra.  They have
tried to exploit approximate isospin and $SU(4)$ symmetries in the nuclear
Hamiltonian and to extend well-known many-body methods in novel ways.   In spite
of all this effort, we know the matrix elements with only limited accuracy.
In this section we review the state of the nuclear-structure
calculations and discuss ways to improve them.  

Most recent attempts to calculate the nuclear matrix elements have been based
on the neutron-proton Quasiparticle Random Phase Approximation (QRPA) or
extensions to it. Of those that haven't, the most prominent are based on the
shell model (SM). While the two methods have much in common --- their
starting point is a Slater determinant of independent nucleons ---
the kinds of correlations they include are complementary.  The QRPA treats a
large fraction of the nucleons as ``active" and allows these nucleons a large
single-particle space to move in.  But RPA correlations are of a specific and
simple type best suited for collective motion.  The shell model, by contrast,
treats a small fraction of the nucleons in a limited single-particle space, but
allows the nucleons there to correlate in arbitrary ways.  That these very
different approaches
yield similar results indicates that both capture most of the important
physics. That, by itself, is encouraging and restricts the possible 
values of nuclear matrix elements considerably.

\subsection{QRPA}

The application of QRPA to $\beta\beta$ decay began  with the realization by
\cite{VZ86} that in the QRPA the neutron-proton ($np$)
particle-particle (i.e.\
pairing-like) interaction, which has little effect on the collective Gamow-Teller
resonance, suppresses \BBt\ rates considerably.  Soon afterward,
\cite{Engel88} and \cite{Tomoda87} demonstrated a
similar though smaller effect on \BBz\ decay.  It was
quickly realized, however, that the QRPA was not designed to handle realistic
$np$ pairing; the calculated half-lives were unnaturally sensitive to the
strength of the pairing interaction.  As a result, the rates of \BB\ decay,
particularly \BBt\ decay, were hard to predict precisely because a
small change in a phenomenological parameter (the strength of $np$
isoscalar pairing) caused a large
change in the lifetimes and eventually the breakdown (called a ``collapse") of
the entire method when the parameter exceeds some critical value.  Most recent
work in the QRPA has aimed at modifying the undesirable aspects of the
method so that its sensitivity to $np$ pairing becomes more realistic.

There has been a large number of attempts to extend the validity of QRPA.
We will not list them here, or discuss them in detail. Comprehensive discussion
can be found in \cite{ELL04}. Earlier and even more complete discussion
of the issues involved can be found in \cite{SuhCiv98}.

Unfortunately, it's not clear which of the extensions of the standard
QRPA is best.  
Some of them
violate an important sum rule for single $\beta$ strength, and studies in
solvable models suggest that the reduced dependence (at least of the 
`renormailzed QRPA (RQRPA)) on
neutron-proton pairing may be spurious resulting from
an artificial
reduction of isoscalar pairing correlations. 
There are not many obvious 
reasons to prefer one of these QRPA extensions or the other.

Recently, one of the methods, RQRPA, was modified even further, so that the single $\beta$
sum rule was restored.
The resulting method, called the ``Fully Renormalized QRPA"
has yet to be applied to \BBz\ decay.  Even more recently, Ref. \cite{Simk03}
raised the issue of nuclear deformation, which has usually been ignored in
QRPA-like treatments of nearly spherical nuclei. (Psuedo-SU(3)-based
truncations have been used to treat it in well-deformed nuclei \cite{Hirsch96}).
The authors of \cite{Simk03} argued that
differences in deformation between the initial and final nuclei can 
have large effects on the \BBt\ half-life. These ideas, too, have not 
yet been applied to \BBz\ decay.

The profusion of RPA-based extensions is both good and bad.  The sheer 
number of methods applied gives us a kind of statistical sample of 
calculations, which could give  an idea of the 
theoretical
uncertainty in the matrix elements.  But the sample may be biased by the
omission of non-RPA correlations in all but a few calculations.  Also,
different calculations are done with different ways of fixing the parameters,
e.g. some insist to reproduce the \BBt\ rate, other are unable
to do that (or do not insist on it), etc. It is not clear, therefore,
that the spread of the calculated values indeed corresponds to the spread of
the possible matrix elements within even just QRPA and its extensions.

\subsection{Shell Model}

The obvious alternative to QRPA, and the current method of choice for nuclear
structure calculations in heavy nuclei where applicable, is the shell model.
It has ability to represent the nuclear wave function to arbitrary accuracy,
provided a large enough model space is used.  This caveat is a huge one,
however.
Current computers allow very large bases (millions of states), but in heavy
nuclei this is still not nearly enough.  Techniques for constructing
``effective'' interactions and operators that give exact results in truncated
model spaces
exist but are hard to implement.  Even in its crude form with relatively small
model spaces and bare operators, however, the shell model seems to offer some advantages
over the QRPA.  Its complicated valence-shell correlations, which the QRPA
omits (though it tries to compensate for them by renormalizating parameters)
apparently affect the \BB\ matrix elements \cite{Caurier96}.

The first modern shell-model calculations of \BB\ decay date from the work
Haxton and Stephenson \cite{HaxSte84} and references therein.   Only a few truly
large-scale shell model calculations have been
performed.  The heavy deformed \BB\ nuclei, $^{238}U$, and 
$^{150}$Nd, for example,
require bases that are too large to expect real accuracy.  Realistic work has
thus been restricted to few nuclei, in particular to
$^{48}$Ca, $^{76}$Ge, and $^{136}$Xe, though less
comprehensive calculations have been carried out in several other nuclei
\cite{SuhCiv98}.

Large spaces challenge us not only through the problem of diagonalizing large
matrices, but also by requiring us to construct a good effective interaction.
The bare nucleon-nucleon interaction needs to be modified in truncated spaces
(this is an issue in the QRPA as well, though a less serious one). 
Currently, effective interactions are
built through a combination of perturbation theory, phenomenology, and
painstaking fitting.  The last of these, in particular, becomes increasingly
difficult when millions of matrix elements are required.

Related to the problem of the effective interaction is the renormalization of
transition operators.  Though the problem of the effective 
Gamow-Teller operator which enters directly into \BBt\ decay, has drawn
some attention,
very little work has been done on the renormalization of the two-body operators
that govern \BBz\ decay.  Shell model 
calculations won't be truly reliable until they address this issue, 
which is connected with deficiencies in the wave function caused by 
neglect of single-particle levels far from the Fermi surface.  Ref. \cite{EngV04} suggests that significant
improvement on the state of the art will be difficult but not 
impossible in the coming years.

\subsection{Constraining Matrix Elements with Other Observables}

The more observables a calculation can reproduce, the more trustworthy it
becomes.  And if the underlying model contains some free parameters, these
observables can fix them.  The renormalization of free parameters can make up
for deficiencies in the model, reducing differences between, e.g., the QRPA
and RQRPA once the parameters of both have been fit to
relevant data.  The more closely an observable resembles \BBz\ decay, the more
relevant it is.

Gamow-Teller distributions, both in the $\beta^-$ and $\beta^+$ directions,
enter indirectly into both kinds of \BB\ decay, and are measurable through
$(p,n)$ and analogous nucleon exchange reactions.   
Ref.\cite{AunSuh98} is particularly careful to
reproduce those transitions as well as possible.  Pion double charge exchange,
in which a $\pi^+$ enters and a $\pi^-$ leaves, involves the transformation of
two neutrons into two protons, like \BB\ decay, but the nuclear operators
responsible aren't the same in the two cases. Perhaps the most relevant
quantity for calibrating calculations of \BBz\ decay is \BBt\ decay, which has
now been measured in 10 different nuclei.

Two recent papers have tried to use \BBt\ decay to fix the strength of $np$
pairing in QRPA-based calculations.  Ref. \cite{Stoica01}
used it only for the
$J^{\pi}=1^+$ channel relevant for \BBt\ decay, leaving the $np$ pairing
strength unrenormalized in other channels.  By contrast, Ref \cite{Rodin03}
renormalized the strength in all channels by the same amount.  The results of
the two procedures were different; the former reference
found that the \BBz\ matrix elements depended significantly on the theoretical
approach, while the later one found almost no dependence on model-space
size, on the form of the nucleon-nucleon interaction, or on whether the QRPA or
RQRPA was used.  The authors argued that fixing the $np$ pairing strength to
\BBt\ rates essentially eliminates uncertainty associated with variations in
QRPA calculations of \BBz\ rates, though they left open the question of how
close to reality the calculated rates were.

The conclusion of \cite{Rodin03} is supported by the work \cite{Muto97}
where the \BBt\ was also used to fix the relevant parameter and essentially
no difference between QRPA, RQRPA, and the new variant developed there was
found. Moreover, when the fact that higher order terms are not included
in \cite{Muto97}, but are included in \cite{Rodin03} (a reduction by $\sim$ 30\%)
are taken into account, these two calculations agree quite well.
Another case is the work \cite{Bobyk01} which uses yet another variant
of the theory and seemingly disagrees drastically with other calculations
for the case of $^{76}$Ge \BBz\ decay. However, in an earlier work
using the same method and parameters \cite{Bobyk99}, it is clear that
the rate of the \BBt\ decay is incorrect by a large factor. When the
relevant parameters are adjusted to give the correct \BBt\ rate, the
result again agrees quite well with \cite{Rodin03}. 

While this question remains open, it is clear that only calculations of
the \BBz\ nuclear matrix elements that also describe other relevant
nuclear properties should be included in the estimate of the uncertainty.

\subsection{Reducing the Uncertainty}

What can be done to improve the situation?  In the near term, improvements can
be made in both QRPA-based and shell-model calculations.  First, existing
calculations should be reexamined to check for consistency.  One important
issue
is the proper value of the axial-vector coupling constant
$g_A$, which is often set to 1 (versus its real value of 1.26) in calculations
of $\beta$ decay and \BBt\
decay to account for the observed quenching of low-energy 
Gamow-Teller strength. What value should one use for
\BBz\ decay, which goes through intermediate states of all multipolarity, not
just $1^+$?  Some authors use $g_A=1$, some $g_A=1.26$, and some $g_A=1$ for
the $1^+$ multipole and 1.26 for the others. (Often authors don't reveal their
prescriptions.)  The second of these choices
appears inconsistent with the treatment of \BBt\ decay.
Since the square of $g_A$ enters
the matrix element, this issue is not totally trivial.
The striking
results of Ref. \cite{Rodin03} suggest that an inconsistent treatment is responsible
for some
of the spread in the calculated nuclear matrix elements.  More and better charge-exchange experiments
would help teach us whether higher-multipole strength is also quenched.

Next, the various
versions of the QRPA should be tested against exact solutions in a solvable
model that is as realistic as possible.  The most realistic used so far are the
$SO(5)$-based model used to study the
QRPA and RQRPA for Fermi \BBt\ decay \cite{Hirsch97}, a
two-level version of that model used in \cite{EngV04} for
the QRPA in
Fermi \BBt\ and \BBz\ decay, and an $SO(8)$-based model
used to test the QRPA and RQRPA for both Fermi and Gamow-Teller \BBt\ decay
in Ref. \cite{Engel97}.  It should be possible to extend the $SO(8)$ model to several
sets of levels and develop techniques for evaluating \BBz\ matrix elements in
it.  All these models, however, leave out spin-orbit splitting, which weakens
the collectivity
of $np$ pairing. Using these models should help to understand the virtues and
deficiencies of various QRPA extensions. 

Along the same lines, we will need to understand the extent to which such
methods can reproduce other observables, and their sensitivity to remaining
uncertainties in their parameters.  A  careful study of the first issue was
made in \cite{AunSuh98}.
These efforts must be
extended.  The work is painstaking, and perhaps not as much fun as concocting
still more variations of the QRPA, but it is crucial if we are to reduce
theoretical uncertainty.  Self-consistent Skyrme HFB+QRPA, applied to
single-$\beta$
decay in \cite{Engel99} and Gamow-Teller resonances
in \cite{Bender02}, may be helpful here; it offers a more general
framework, in
principal anyway, for addressing the variability of calculated matrix elements.
Solvable models can be useful here too, because they can sometimes supply
synthetic data to which parameters can be adjusted (as in \cite{EngV04}).

The best existing shell-model calculation produces smaller matrix elements
than most QRPA calculations.  Computer speed and memory is now at the point
where the state of the shell-model art can be improved.  The calculation of the
\BB\ decay of $^{76}$Ge in \cite{Caurier96}used the
$f_{5/2}p_{3/2}p_{1/2}g_{9/2}$ model space, allowing up to 8 
particles (out of a
maximum of 14) into the $g_{9/2}$ level.  Nowadays, with the help of the
factorization method (see \cite{Papen03,Papen03a}), an
accurate approximation to full shell-model
calculations, we should be able to fully occupy the $g_{9/2}$ level, and
perhaps include the $g_{7/2}$ and $f_{7/2}$
levels (though those complicate things by introducing spurious center-of-mass
motion).
In addition, one can try through diagrammatic perturbation theory to
construct effective \BBz\ operators for the model space that are 
consistent with
the effective interaction.  Though perturbation theory has 
convergence problems,
the procedure should at least give us an idea of the uncertainty in the
final answers, perhaps also indicating whether result obtained from 
the ``bare''
operators is too large or too small.  Research into effective operators has
been revived in
recent years \cite{Hax00} and we can hope to improve on 
diagrammatic perturbation
theory.   One minor source of uncertainty connected with renormalization 
(which also affects the QRPA) is short-range two-nucleon correlations,
currently treated phenomenologically,
following \cite{Miller76}.

In short, much can be done and we would be well served by coordinated
attacks on these problems.  There are relatively few theorists working
in \BB\ decay, and their efforts have been fragmented.  More
collaborations, postdoctoral and Ph.D\ projects, meetings, etc.,
would make progress faster.  {\it There is reason to be hopeful that
the uncertainty will be reduced.}  The shell-model matrix element may be too
small because it does not include any particles outside the $fp+g_{9/2}$-shell.  These
particles, as shown by QRPA calculations, make the matrix element larger.  We
suspect that the results of a better shell-model calculation will be closer
than the best current one to the QRPA results and that, as noted above,
the spread in those
results can be reduced.  Finally, other nuclei may be
more amenable to a good shell-model calculation than Ge.  $^{136}$Xe has 82
neutrons (a magic number) making it a particularly good candidate.

\section{IV. Cosmology and Neutrino Mass}
Neutrinos play an important role in
astrophysics and cosmology (see also section IIID).  In cosmology, relic neutrinos
may constitute an important fraction of the hot dark
matter (HDM) influencing the evolution of large scale structures
(LSS)\cite{cosmo_review}.
The imprint of neutrino HDM on LSS
evolution is quite distinct from other dark matter candidates such
as supersymmetric particles, which act as non-relativistic or Cold
Dark Matter (CDM). Cosmological models of structure formation
strongly depend on the relative amounts of CDM and HDM in the
universe, hence a determination of the neutrino contribution
$\Omega_{\nu}$ to the total dark matter content $\Omega$ of the
universe is important for our understanding of structure formation
\cite{cosmo_review}.
Such an investigation is a strong motivation
for the next-generation terrestrial neutrino mass experiments.
For example, a single $\beta$-decay experiment
with a neutrino mass sensitivity of 0.2 eV (\,90\,\%CL.) 
will be sensitive to a neutrino contribution
$\Omega_\nu$\,=\,0.015. Such measurements would either
significantly constrain or fix the role of HDM in
structure formation by a model independent method.

Two areas of astrophysics where an understanding of
neutrino mass is important are supernova dynamics and cosmic ray
physics. In the area of cosmic rays, a new model has
been proposed which aims at explaining the origin of ultra high
energy (UHE) cosmic rays. This so-called Z-burst model would
require relic neutrino masses within the sensitivity range of
next generation neutrino mass experimments.

\subsection{Cosmological studies and neutrinos}

The contribution $\Omega_{\nu}$ of relic neutrinos to the total
density $\Omega$ can be probed by a combination of precise
measurements of the temperature fluctuations of the 
cosmic microwave background radiation and with
measurements of the matter fluctuation at large scales by high
statistics galaxy redshift surveys. The results for the mass
density of neutrinos can then be transformed into neutrino mass
results based on the calculated relic neutrino density of
112\,$\nu$'s/cm$^3$ in the framework of the canonical standard
cosmological model. (Detailed information on these neutrinos
and cosmology can be found in the Astrophysics and Cosmology 
working group report.)

\subsubsection{Results from cosmological studies}

Early cosmological bounds on $\Sigma$m$_{\nu}$,
based on pre-WMAP/SDSS/ 2dFGRS data, gave upper limits in the
range of 3\,--\,6 eV, comparable to the present laboratory limit
from tritium $\beta$-decay. 
Recent cosmological studies are based on different combinations of
the high quality data from WMAP, small scale high resolution CMBR
experiments, and the two large galaxy redhift surveys 2dFGRS and
SDSS. In some cases additional structure information from
Lyman-$\alpha$ data or X-ray luminosities from clusters have been
added. 

Table \ref{cosmo_results} presents a summary of either upper limits or best values for neutrino
masses derived from recent cosmological studies.
The table shows the considerable spread in the results published
recently. This is due to several generic difficulties associated with
cosmological studies. First, these studies suffer from the problem
of parameter degeneracy \cite{hannestad}. Different combinations
of cosmological parameter can describe the LSS and CMBR data
equally well, so additional information is required to break the
degeneracy. The errors
associated with these input parameters also imply that the $\nu$-mass
results crucially depend on the \textit{priors} for these
parameters. Different priors for cosmological parameters result in
limits for the sum of neutrino masses $\Sigma$m$_{\nu}$ which
differ by factors of 2 or more. 

The strong dependence of cosmological neutrino mass results can be
illustrated by comparing the strongest upper limit on m$_{\nu}$
reported in the literature, the WMAP upper limit \cite{wmap} of
m$_{\nu}$\,$<$\,0.27\,eV (95\,\%\,CL.), with tentative evidence
for non-zero neutrino masses reported by Allan {\it et al.} \cite{nonzero}. Both
analyses use substantially identical sets of input parameters, most
notably the WMAP and 2dFGRS data. However, while the WMAP
authors\cite{wmap} use additional Lyman-$\alpha$ data for their analysis,
Allan {\it et al.}  \cite{nonzero} use X-ray luminosity functions (XLF)
of galaxy clusters obtained with the orbiting Chandra X-ray
telescope. This small change of input data transforms an upper
limit into evidence for non-zero masses with
$\Sigma$m$_{\nu}$=0.56\,eV as best fit value. It is interesting to
note that both the use of Lyman-$\alpha$ data as well as the use of
the measured XLF have been criticized in the literature. This
clearly underlines that cosmological studies, as impressive as
they are with regard to the determination of cosmological
parameters, still yield \textit{model-dependent} results for
neutrino masses.

Further concerns with cosmological neutrino mass studies are
associated with systematic errors. LSS data from galaxy surveys
suffer from the problem of biasing (i.e. to what extent does the
galaxy distributing trace the distribution of cold dark matter and
dark baryons), possible redshift-space distortions and selection
effects. A model-independent input of the value of the neutrino
mass to cosmological studies would thus be especially important
for future high precision cosmological studies. A laboratory
measurement of m$_{\nu}$ could break the existing degeneracies
between $\Omega_{\nu}$ and other cosmological parameters, and thus
help to provide a better picture of large scale structure
evolution.

\begin{table}
  \centering
\begin{tabular}{|l|c|c|c|c|c|r|}
  \hline  \hline

   author       & WMAP & CMB$_{hi-l}$ & SDSS & 2dF & other data & $\Sigma$\,m$_{\nu}$\,[eV]\\
   \hline
   Bar'03 \cite{barger}    & x & x & x & x & h\,(HST)           & $<$\,0.75 \\ \hline
   Teg'03 \cite{sdss}      & x & x & x &   & SNIa               & $<$1.7 \\ \hline   ASB'03 \cite{nonzero}   & x & x &   & x & XLF                & =\,0.36-1.03 \\ \hline
   WMAP   \cite{wmap}      & x & x &   & x & Ly$\alpha$\,,\,h\,(HST) & $<$\,0.7 \\ \hline
   Bla'03 \cite{blanchard} & x &   &   & x & $\Omega_m$=1       &   =\,2.4 \\ \hline
   Han'03 \cite{hannestad} & x & x &   & x & h\,(HST),\,SNIa    & $<$\,1.01 \\ \hline
   Han'03 \cite{hannestad} & x & x &   & x &                    & $<$\,1.2 \\ \hline
   Han'03 \cite{hannestad} & x &   &   & x &                    & $<$\,2.12 \\ \hline
  \hline

\end{tabular}
  \caption{Survey of neutrino mass results obtained from the most recent cosmological studies.
   For each study the specific set of input data is listed: WMAP angular and/or polarisation data, high resolution
   CMBR experiments like CBI etc, LSS data from the SDSS and the 2dFGRS galaxy surveys, and other
   data. }
  \label{cosmo_results}
\end{table}

\subsection{Beyond the 'concordance' model}

The results of the cosmological studies presented above in table
\ref{cosmo_results} have been obtained within the 'canonical
standard cosmological model'. In the following we briefly list
studies which are outside the present so-called 'concordance' flat
$\Lambda$CDM models.

 In \cite{blanchard} the authors argue that
cosmological data can be fitted equally well in the framework of
an Einstein-de Sitter universe with with \emph{zero }cosmological
constant $\Lambda$, albeit at the expense of requiring a very low
value for the Hubble constant of H$_{0}$\,$\simeq$\,46 km/s/Mpc.
The authors  claim that CMB and LSS data
seem to imply the existence of non-cold dark matter component.

I n addition,several authors \cite{paes03} have speculated about the
interesting coincidence between the smallness of neutrino masses
(m$_{\nu}$\,$<$\,1\,eV) and the deduced vacuum density $\rho_V$
$\approx$ (\,10$^{-3}$ eV\,)$^4$ responsible for the cosmological
dark energy $\Lambda$. 
In this framework, the masses of the active $\nu$-species vary like
the inverse of the $\nu$-density. Accordingly, the neutrino mass
could be of order of 1~eV today.

Another 'coincidence' problem ($\Omega_{\Lambda}$ = $\Omega_m$)
has been used by M Tegmark et al. \cite{anthropic} to investigate
the possibility that m$_{\nu}$ is a stochastic variable, which is
randomized during inflation. 

Yet another possible modification to the neutrino sector of the
'concordance' model has been brought up by the authors of
\cite{v_less}. They assume that neutrinos have small extra scalar
or pseudoscalar interactions. Should neutrinos possess even only
tiny couplings of the order of 10$^{-5}$ to hypothetical scalar
$\phi$ bosons, the neutrino density $\Omega_{\nu}$ could be
affected strongly by annihilation processes of the type
$\nu$\,$\nu$\,$\leftrightarrow$\,$\phi$ $\phi$. In this scenario
neutrinos would not decouple from matter at T\,=\,1\,MeV, but
would stay in thermal equilibrium until much later times
(T\,=\,1eV), which would inhibit free streaming.

Relic neutrinos have not been detected at present, and will likely
not be detected in the nearer future due to their very low energy
in the $\mu$eV range. For this reason, the 'neutrinoless universe'
scenario can only be tested by comparing the model-independent
neutrino mass result from single and double beta decay experiments
along with cosmological data. 

\subsection{Future Perspectives}

At present, the cosmological studies of m$_{\nu}$ are still
model-dependent due to systematic effects such as biasing,
parameter degeneracy, possible selection effects, possible
contributions from non-linear effects and the strong influence of
priors on the neutrino mass results. Future high precision studies
aim at strongly reducing these systematic effects.

In the field of CMBR experiments, the new promising technique of
studying distortions of the CMBR temperature and polarization maps
induced by gravitational lensing has been proposed \cite{lensing}.
This technique is sensitive to changes of structure evolution at
late times due to massive neutrinos and thus could break the
degeneracy between neutrino mass, equation-of-state of the dark
energy and the primordial power spectrum amplitude. In
\cite{lensing} the sensitivity of this method is estimated to come
down to $\Sigma$m$_{\nu}$=0.3 eV. This is of the same order as
what is expected in \cite{eisenstein} for the $\nu$-mass
sensitivity (m$_{\nu}$=0.14 eV) from the Planck satellite
scheduled to start operations in 2008.

These investigations will be complemented by future deep galaxy
surveys extending out to redshift parameters z=2 (DEEP2,
VLT-Virmos) as well as dedicated studies of lensing effects on
galaxy clusters (LSST, Large Synoptic Survey
Telescope).

\section{V. Experimental Prospects for $\beta$ Decay}

\subsection {Motivation for absolute measurements}

The SM of particle physics describes present experimental data up
to the electroweak scale, but does not explain the
observed pattern of the fermion masses or the mixing among the
fermion generations. In addition it explicitly assumes neutrinos
are massless and offers no explanation for the observed 
$\nu$-masses and $\nu$-mixing. 

There are many theories beyond the Standard Model, which explore
the origins of neutrino masses and mixing. In these theories,
which often work within the framework of Supersymmetry, neutrinos
naturally acquire mass. A large group of models makes use of the
so-called see-saw effect to generate neutrino masses.
Other classes of theories are based on completely
different possible origins of neutrino masses, such as radiative
corrections arising from an extended Higgs sector.
 As neutrino masses are much smaller than the
masses of the other fermions, the knowledge of the absolute values
of neutrino masses is crucial for our understanding of the fermion
masses in general. Recently it has been pointed out  that the {\it
absolute mass scale of neutrinos} may be even more significant and
straightforward for the fundamental theory of fermion masses than
the determination of the neutrino mixing angles and CP-violating
phases \cite{smirnov}. It will likely be the absolute mass
scale of neutrinos which will determine the scale of new physics.

\subsubsection {Introduction (related to kinematical experimental techniques)}

Two kinematical experimental techniques that probe neutrino mass directly,
are precise observations of decay
kinematics in nuclear or particle decay processes and the
utilization of time of flight measurements in
the detection of supernova neutrinos incident on terrestrial neutrino
detectors.  The decay measurements, which
are based on purely kinematical observables,
have essentially no reliance on theoretical assumptions about neutrino
properties and hence offer a completely model independent probe of
absolute neutrino mass.  

Examples of decay
measurements include studies of $\beta$ -decay spectra shapes,
measurements of muon momentum in pion decay,
and invariant mass studies of multi-particle semileptonic decays of
the $\tau$. However, the recent measurement of large mixing angles in the
lepton sector gives a clear advantage to $\beta$ -decay spectra shape
measurements, since these experiments probe all three neutrino
mass eigenstates and are approaching the sub-eV range os sensitivity. 
These next generation  $\beta$ -decay measurements will complement other
laboratory and cosmological
methods to investigate neutrino masses. The combination and
comparison of results from $\beta$-decay, 0$\nu\beta\beta$
and cosmological studies will be essential for our understanding
of the role of neutrinos in our physical world, both at the
Micro- and Macro world.


In contrast to \BBz\ experiments, kinematic investigations of
the neutrino mass do not rely on further assumptions on the
neutrino mass type, Majorana or Dirac. Such kinematic
experiments can be classified into two categories both making use
of the relativistic energy momentum relation $E^2 = p^2 c^2 + m^2
c^4$ as well as of energy and momentum conservation.

\subsubsection{Neutrino time-of-flight studies}

The narrow time signal of a supernova (SN) neutrino burst of less
than 10~s in combination with the very long-baseline between
source and detector of several kpc would allow the investigation
of small ToF
effects resulting from small $\nu$-masses.
Supernova ToF studies are based on the observation of the
energy-dependent time delay of massive neutrinos relative to
massless neutrinos. This method provides an experimental
sensitivity for the rest masses of \nue, \numu and \nutau of a
few tens of eV. This sensitivity can be pushed into the few-eV
range, if additional assumptions on the time evolution of the
$\nu$-burst are being made. In this case the $\nu$-mass
sensitivity becomes model-dependent, however.

Recently, new methods have been proposed which do not rely on
details of the $\nu$-burst timing which might yield
sensitivity in the few-eV range for \numu\ and \nutau .
The two most promising techniques are: a) the measurement of the
abrupt termination of the SN-$\nu$ signal due to the early
formation of a black hole \cite{sn_bh1,sn_bh2} and b) the
correlation of SN $\nu$-signals with independent signals from
gravitational wave experiments \cite{sn_grav1}.  
Except for exceedingly close by
supernova, $<$\,1\,kpc , the
sensitivity from the first technique to neutrino mass would still only 
be at the level of a few eV.  Estimates for the sensitivity to
the second method are also expected to also be only at the eV level.

If future measurements were to reveal a SN $\nu$-pulse which
is terminated abruptly after a few seconds, the information on the
$\nu$-mass provided by next-generation terrestrial experiments
would likely be used to help further our understanding of supernovae dynamics.

\subsubsection{Weak particle decay processes}
The investigation of the kinematics of weak decays is based on the
measurement of the charged decay products of weak decays. For the
masses of \numu\ and \nutau\ the measurement of pion decays into
muons and \numu\ at PSI and the investigation of $\tau$-decays
into 5 pions and \nutau\ at LEP have yielded the upper limits:
\begin{eqnarray*}
  m(\numu) &<& 190~\mbox{ keV~~~~~~ at~ 90\, \%~ confidence ~\cite{pdg}} \\
  m(\nutau) &<& 18.2~\mbox{ MeV~~~~ at~ 95\, \%~ confidence ~\cite{pdg}}
\end{eqnarray*}
Both limits are much larger than those attainable in beta decay spectra studies.

\subsection{\bdec}\label{sstrit}

The technique for detecting neutrino mass in beta decay is
essentially to search for a distortion
in the shape of the beta spectrum in the endpoint energy region.
The most sensitive searches for the electron neutrino mass
up to now are based on the investigation of the electron spectrum
of tritium \bdec\

\be
  ^3{\rm H} \rightarrow  ^3{\rm He}^+ + \el  + \nueb \quad {\rm .}
\ee

The electron energy spectrum of tritium \bdec\ for a neutrino with
mass $m_\nu$ is given by
\begin{equation}
{dN \over dE} = C \times F(Z,E) p E(E_0-E) [(E_0-E)^2-m_\nu^2]^{1
\over 2} \Theta (E_0-E-m_\nu), \label{mother}
\end{equation}
where $E$ denotes the electron energy, $p$ is the electron
momentum,  $E_0$ corresponds to the total decay energy, $F(Z,E)$
is the Fermi function, taking into account the Coulomb interaction
of the outgoing electron in the final state, the stepfunction
$\Theta (E_0-E-m_\nu)$ ensures energy conservation, and $C$ is
given by \be C=G_F^2 {m_e^5  \over 2 \pi^3} \cos^2 \theta_C |M|^2
~. \label{re} \ee Here $G_F$ is the Fermi constant, $\theta_C$ is
the Cabibbo angle, $m_e$ the mass of the electron and $M$ is the
nuclear matrix element. As both  $M$ and $F(Z,E)$ are independent
of $m_\nu$, the dependence of the spectral shape on $m_\nu$ is
given by the phase space factor only. In addition, the bound on
the neutrino mass from tritium \bdec\ is independent of whether
the electron neutrino is a Majorana or a Dirac particle.

\begin{figure}
  \includegraphics[width=5.0 in]{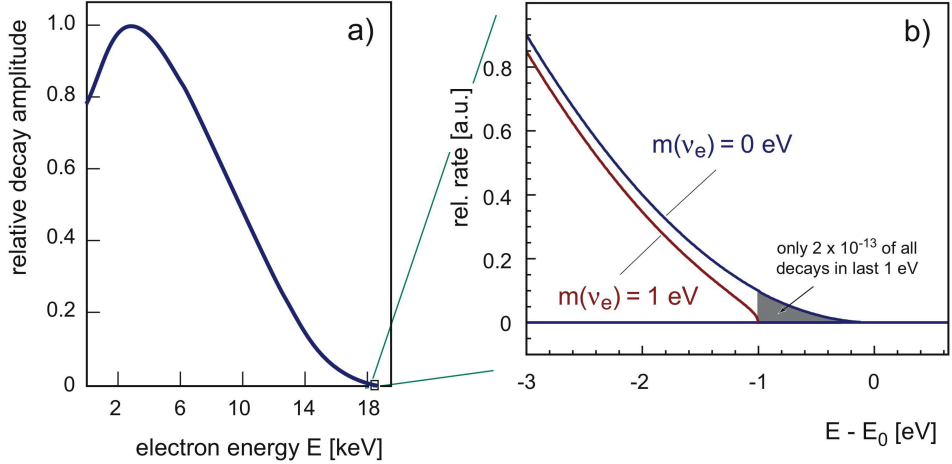}
  \caption{ The electron energy spectrum of tritium \bdec : (a) complete and (b) narrow
   region around endpoint \ezero . The \bspec\ is shown for neutrino
   masses of 0 and 1~eV.
  }
  \label{fig_betaspec}
  \end{figure}

The signature of an electron neutrino with a mass of
m(\nue\,)=1\,eV is shown in Fig.~\ref{fig_betaspec}, in comparison with the
undistorted $\beta$ spectrum of a massless \nue\,. The spectral
distortion is statistically significant only in a region close to
the $\beta$ endpoint. This is due to the rapidly rising count rate
below the endpoint ${\rm dN/dE} \propto (E_0 - E)^2$. Therefore,
only a very narrow region close to the endpoint $E_0$ is analyzed.
Note that in fitting the measured
shape to a calculated shape the functional
form depends on $m_\nu^{2}$ and is not defined for $m_\nu^{2} < 0$.

One immediate difficulty apparent in the figure is that there are
very few decays in the
region of interest, only $2 \cdot 10^{-13}$ of the total rate
is in the last eV. A related problem is that
one must carefully  eliminate or minimize backgrounds. 
In addition subtle effects such as instrumental
resolution, energy loss of the electrons, and
atomic or molecular physics excitations during the decay can cause shape
distortions that are of similar size to the
effect of nonzero neutrino mass but of opposite sign. 
Hence tritium \bdec\ experiments with
high neutrino mass sensitivity require a huge luminosity combined
with very high energy resolution.
Furthermore, a precise
determination of a value or limit for
 $m_\nu$   requires complete and accurate
understanding of all systematic effects that can
alter the shape of the spectrum.

Apart from offering a low endpoint energy $E_0$ and a moderate
half life of 12.3\,y, tritium has further advantages as $\beta$
emitter in $\nu$ mass investigations:
\begin{enumerate}
\item the hydrogen isotope tritium and its daughter, the
$^3$He$^+$ ion,
      have a simple electron shell configuration.  Atomic corrections
       for the \bdec ing atom -or molecule- and corrections due to the interaction
      of the outgoing \belec\ with the tritium source can be calculated in a simple and
      straightforward manner
\item The tritium \bdec\ is a super-allowed nuclear transition.
Therefore, no corrections
      from the nuclear transition matrix elements $M$ have to be taken into account.
\end{enumerate}
The combination of all these features makes tritium an almost
ideal $\beta$ emitter for neutrino mass investigations.

\subsection{Current tritium $\beta$-decay results}

The Mainz and Troitsk groups have
set the most precise
limits on the electron antineutrino mass.
Both experiments utilize novel magnetic solenoidal retarding electrostatic spectrometers
which measure an
integral beta spectrum,
integrating all energies above
the acceptance energy of the spectrometer.  
In their measurements, the Mainz group utilized a frozen molecular tritium source.  Their
result \cite{Weinheimer03} is:
\begin{equation}
m_{\nu_{e}}^{2} = -1.2 \pm 2.2 \pm 2.1 ~eV^{2},\nonumber\\
\end{equation}
which yields a limit of:
\begin{equation}
m_{\nu_{e}} <2.2 ~eV  ~~(95 \%CL).\\
\end{equation}
This result is based on
data that has passed several systematic and consistency checks.
 The Troitsk group\cite{lob00,Lobashev04} developed a gaseous molecular tritium source and
has also published a limit similar to that of the Mainz group of 
\begin{equation}
m_{\nu_{e}}^{2} = -2.3 \pm 2.5 \pm 2.0 ~eV^{2},\nonumber\\
\end{equation}
with a limit of:
\begin{equation}
m_{\nu_{e}} <2.1 ~eV  \nonumber ~~(95 \%CL).\\
\end{equation}
However, they must include a not well understood
step function near the endpoint in order to produce such a limit.

\subsection{Next generation experiments}
\subsubsection{The {\bf KA}rlsruhe {\bf TRI}tium {\bf N}eutrino project
(KATRIN) experiment}
The {\bf KA}rlsruhe {\bf TRI}tium {\bf N}eutrino project
(KATRIN) experiment
is a next-generation tritium $\beta$-decay experiment
designed to measure the mass of the neutrino with
sub-eV sensitivity\cite{katrinloi}.  KATRIN utilizes a windowless gaseous molecular tritium source and a 10 m diameter magnetic solenoidial retarding electrostatic type spectrometer. The experiment will be constructed adjacent to the Tritium Laboratory Karlsruhue at the Forschungszentrum Karlsruhe (FZK).  The collaboration membership includes a significant number of collaborators from the most recent successful tritium $\beta$-decay experiments, including Mainz and Troitsk, as well as the Washington group which developed the original LANL gaseous based tritium source. The current schedule calls for for KATRIN to become operational during 2008.  The experiment expects to achieve an order of magnitude
more sensitivity than the best $\beta$-decay measurements carried out to date.
(Note, since the shape effects are proportional to $m_\nu^{2}$ this implies a factor of 100 increase in sensitivity.)
Based on the current design, KATRIN expects after three years of running to reach a sensitivity to
neutrino mass of 0.20 eV (90\% CL) and would hence be able to observe
a neutrino mass with a mass of 0.35eV at the 5 sigma significance level.
The experiment has recently received a firm funding commitment from the German funding ministry. 

\subsubsection{ NEXTEX (Neutrino Experiment at Texas)}
The {\bf N}eutrino {\bf EX}periment at {\bf TEX}as is an experiment that has been
under construction for more than 10 years with the goal of measuring the electron 
antineutrino mass to 0.5 eV (lower limit at 3$\sigma$). Collaborating institutions 
are The University of Texas at Austin, Nebraska, Brandeis, Michigan Technological 
Institute, Pomona College, and Southwestern University.  The experiment uses the 
end point method for gaseous tritium decay, and the apparatus consists of three 
serial electrostatic differential analyzers.  Presently the experiment is funded
by the NSF to demonstrate proof of principle.  Construction completion and final
approval by the site committee is expected by the collaborators within 1 year. 
The experiment will run for 3 years and the anticipated funding request would be \$3.9M.

\subsubsection{Other approaches to \bdec}

A different approach to directly measure the electron neutrino
mass is the use of cryogenic bolometers. In this case, the $\beta$
source can be identical to the \belec\ spectrometer. This new
technique has been applied to the isotope $^{187}{\rm Re}$, which
has a 7 times lower endpoint energy than tritium\cite{Milano,Genoa}. 
Current
microcalorimeters reach an energy resolution of $\Delta E \sim
5$\,eV for short-term measurements and yield an upper limit of
$m(\nue)<15$\,eV\ \cite{Milano,Genoa}. To further improve the statistical
accuracy, the principle of integration of active source and
detector requires the operation of large arrays of
microcalorimeters. The sensitivity on the neutrino mass
in the next 5-7 years is expected to reach below the few eV level\cite{Milano}.

\section{VI. Experimental Prospects for \BB}

\begin{figure}
\vspace{9pt}
\begin{center}
\includegraphics[width=3. in]{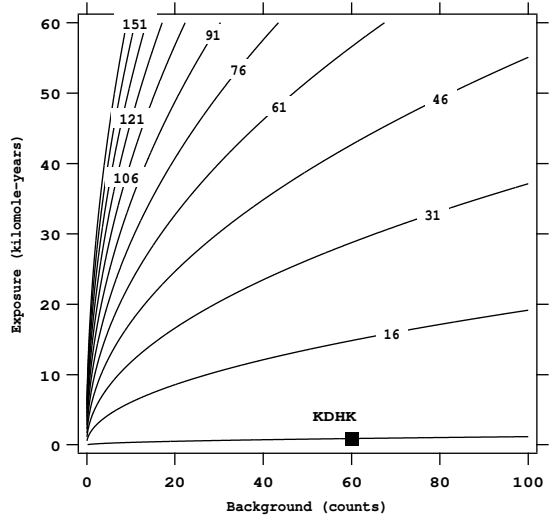}
\end{center}
\caption{This contour plot shows the half life, in units of $10^{25}$ 
y, for a peak of $5\sigma$ significance
for a given exposure and background. The KDHK point is shown.}
\label{fig:sensitivity}
\end{figure}

   If an experiment observes \BBz\ it will have profound physics 
implications. Such an
extraordinary claim will require extraordinary evidence.
The recent claim\cite{KLA04}
for an observation of \BBz\ has been controversial (See discussion 
below). Also previous ``false peaks"
in \BB\ spectra have appeared near a \BBz\ endpoint energy (see 
discussion in \cite{MOE94}, page 273).
One must ask the question: What
evidence is required to convincingly demonstrate that \BBz\ has 
been observed?
Low-statistical-significance peaks ($\approx 2\sigma$)
have faded with additional data, so one must require strong 
statistical significance
(perhaps 5$\sigma$). (See Fig. \ref{fig:sensitivity}.) This will 
require a large signal-to-noise ratio that will most likely be
accomplished by an ultra-low-background experiment whose source is 
its detector. Such experiments
are usually calorimetric and provide little information beyond just 
the energy measurement.

How does an experiment demonstrate that an observed peak 
is actually due to \BB\ decay
and not some unknown radioactivity?
Additional information beyond just an energy measurement may be 
required. For example,
although there is some uncertainty
associated with the matrix elements, it is not so large that a 
comparison of measured rates in
two different isotopes could not be used to demonstrate consistency 
with the Majorana-neutrino hypothesis.
Alternatively, experiments that provide an additional handle on the 
signal, for example by measuring a variety of kinematical
variables, demonstrating that 2 electrons are present in the final 
state, observing the $\gamma$ rays associated
with an excited state, or identifying the daughter nucleus, may 
lend further credibility to a claim.
Experiments that provide this extra handle may require a 
significantly more complicated
apparatus and therefore face additional challenges.

The exciting aspect of \BB\ research today is that many proposed 
experiments intend to reach a Majorana
mass sensitivity of $\sqrt{\delta m_{\rm atm}^{2}}$. Several 
different isotopes and experimental
techniques are being pursued actively and many of the programs look viable.
In this section we describe the current situation in experimental 
\BB\ decay .

\subsection{Results to date}
Table \ref{tab:ZeroNuResults} lists the recent \BBz\ results. The 
best limits to date come
from the enriched Ge experiments. The two experiments had comparable
results although the Heidelberg-Moscow result was marginally 
better. The \Tz\ limits near $2 \times 10^{25}$ y results in a \mee\
limit near 300 meV, with an uncertainty of about a factor of 3 
because of the uncertainty
in \Mz. One recent paper\cite{ZDE02} 
performed a joint analysis of the two experiments and
found \Tz\ $> 2.5 \times 10^{25}$ y.

Most of the results listed in Table \ref{tab:ZeroNuResults} are at least
a few years old. The obvious exceptions to this are the Te and Cd results.
CUORICINO continues to collect data.

\begin{table}
\caption{\protect  A summary of the recent \BBz\ results. The \mee\ 
limits are those
deduced by the authors. All limits are at 90\% confidence level 
unless otherwise indicated. The columns providing the
exposure and background are based on arithmetic done by the authors 
of this paper, who take responsibility for any
errors in interpreting data from the original sources.}
\label{tab:ZeroNuResults}
\begin{center}
\begin{tabular}{lcclc}  \hline\hline
Isotope               & Exposure            & Background  & Half-Life & \mee\    \\
                       & (kmole-y)           &  (counts)   & Limit (y)                  & (meV)    \\ \hline
        $^{48}$Ca      & $5\times 10^{-5}$   &   0         & $>1.4 \times 10^{22}$       &  $<7200-44700$\cite{OGA04}               \\
        $^{76}$Ge      &   0.467             &   21        & $>1.9 \times 10^{25}$       &  $<350$\cite{KLA01}              \\
        $^{76}$Ge      &   0.117             &   3.5       & $>1.6 \times 10^{25}$       &  $<330-1350$\cite{AAL02}               \\
        $^{76}$Ge      &   0.943             &   61        & $=1.2 \times 10^{25}$       &  $=440$\cite{KLA04}               \\
        $^{82}$Se      & $7\times 10^{-5}$   &   0         & $>2.7 \times 10^{22}$(68\%) &  $<5000$\cite{ELL92}               \\
        $^{100}$Mo     & $5\times 10^{-4}$   &   4         & $>5.5 \times 10^{22}$       &  $<2100$\cite{EJI01}              \\
        $^{116}$Cd     & $1\times 10^{-3}$   & 14          & $>1.7 \times 10^{23}$       &  $<1700$\cite{DAN03}              \\
        $^{128}$Te     & Geochem.            &   NA        & $>7.7 \times 10^{24}$       &  $<1100-1500$\cite{BER93}              \\
        $^{130}$Te     & 0.025               &   5         & $>5.5 \times 10^{23}$       &  $<370-1900$\cite{ARN04}              \\
        $^{136}$Xe     & $7\times 10^{-3}$   &   16        & $>4.4 \times 10^{23}$       &  $<1800-5200$\cite{LUE98}              \\
        $^{150}$Nd     & $6\times 10^{-5}$   &   0         & $>1.2 \times 10^{21}$       &  $<3000$\cite{DES97}              \\ \hline
\end{tabular}
\end{center}
\end{table}

\subsubsection{A claim for the observation of \BBz}

In early 2002, a claim for the observation of \BBz\ was published 
(Klapdor-Kleingrothaus \etal  2002a).
The paper made a poor case for the claim and drew strong criticism\cite{AAL02a,FER02,ZDE02}.
The initial response to 
the criticism was emotional\cite{KLA02b}. In addition, one of the original 
co-authors wrote a
separate reply\cite{HAR01} that mostly defended the claim yet 
acknowledged some significant
difficulty with the analysis. This author's name doesn't appear on 
later papers.
More recently, however, supporting evidence for the claim
has been presented and we recommend the reader study Ref. \cite{KLA02c}
for
a good discussion of the initial evidence and Ref. \cite{KLA04}  for the
most recent data analysis. Importantly, this later paper includes 
additional data and therefore an
increase in the statistics of the claim. In this subsection we 
summarize the current situation.
(We use the shorthand KDHK to refer to the collection of papers 
supporting the claim.)

\begin{figure}
\vspace{9pt}
\begin{center}
\includegraphics[width=3. in]{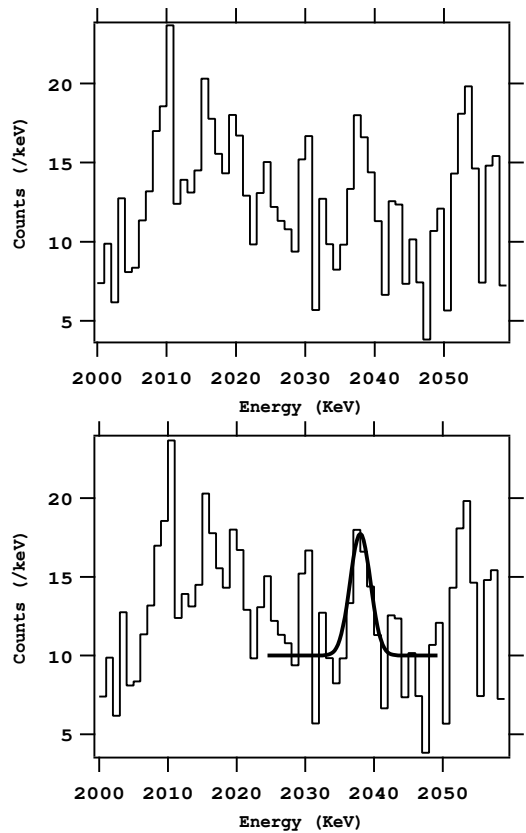}
\end{center}
\caption{The spectrum from the Heidelberg-Moscow experiment upon 
which the claim for \BBz\ is based.
The data in the two panels are identical. The lower panel has a 
Gaussian curve to indicate
the strength of the claimed \BBz\
peak.}
\label{fig:2}
\end{figure}

Figure \ref{fig:2} shows the spectrum corresponding to 71.7 kg-y 
of data from the Heidelberg-Moscow experiment
between 2000 and 2060 keV\cite{KLA04}.
This spectrum is shown here to assist the casual reader in understanding the
issues. However, the critical reader is encouraged to read the 
papers listed in the references as the
authors analyze several variations of this data using different techniques.
The fit about the
expected \BBz\ peak energy yields 28.75 $\pm$ 6.86 counts assigned 
to \BBz. The paper claims
a significance of approxmately $4\sigma$  for the peak, where the 
precise significance value depends on
the details of the analysis.
The corresponding best-fit lifetime, \Tz\ = 1.19 $\times$ 10$^{25}$ years\cite{KLA04},
leads to a \mee\ of 440 meV 
with the matrix element
calculation of Ref. \cite{STA90} chosen by the authors.

In the region between 2000 and 2100 keV, the KDHK analysis of 2002 
found a total of 7 peaks. Four of these
were attributed to $^{214}$Bi (2011, 2017, 2022, 2053 keV), one 
was attributed to \BBz\ decay (2039 keV),
and two were unidentified (2066 and 2075 keV). The KDHK analysis 
of 2004 does not discuss the spectrum
above 2060 in detail. An additional possible feature may also be present
near 2030 keV. A study\cite{KLA03}  
comparing simulation to calibration
with $^{214}$Bi demonstrates that if
the location of the Bi is known, the spectrum can be calculated. 
Furthermore, the relative strengths of
the strong Bi lines at 609, 1764 and 2204 keV can be used to 
determine the location of the activity.
Because the results of summing effects depend on the proximity of 
the activity,
its location is critical for the simulation of the weak peaks near 
the \BBz\ endpoint. The study also shows that the spectrum
can't be naively estimated, as was done in Ref. \cite{AAL02a}. 
In fact, Table VII in Klapdor-Kleingrothaus
(2002c) finds, even with a careful simulation, that the expected 
strengths of the $^{214}$Bi peaks in the 2000-2100 keV
region are not predicted well by scaling to the strong peaks. That
is, the measured intensities of the weak peaks are difficult to
simulate without knowing the exact location of the activity. 
Furthermore, the deduced strengths of the weak lines
are more intense than  expected
by scaling from the strong peaks, even though the activity 
location is chosen to best
describe the relative intensities of the strong peaks.

Double-beta decay produces two electrons that have a short range 
in solid Ge. Therefore, the energy deposit
is inherently localized. Background process, such as the $\gamma$ 
rays from $^{214}$Bi, tend to produce
multiple energy deposits. The pulse waveform can be analyzed to 
distinguish single site events (SSE) from
multiple site events. Such an analysis by KDHK\cite{KLA03a,KLA04} 
 tends to indicate that the 
Bi
lines and the unidentified lines behave as multiple site events, 
whereas the \BBz\ candidate events
behave as SSE. Note, however, that the statistics are still poor 
for the experimental lines and this conclusion
has a large uncertainty.
Nonetheless, this feature of the data is very intriguing and 
clearly a strength of the
KDHK analysis.

An analysis by Zdesenko \etal\cite{ZDE02} points out the strong 
dependence of the result on the choice of
the window width of the earlier 2002 analysis. The KDHK analysis 
argues that a small window is required because of the neighboring
background lines. Even so, their Monte Carlo analysis shows that 
the result becomes less stable for
small windows (see Fig. 9 in Ref. \cite{KLA02c}). 
Zdesenko \etal\cite{ZDE02} also
remind us that the significance of a signal is overestimated when 
the regions used to estimate the
background are comparable to the region used to determine the 
signal\cite{NAR00}. The report of Ref. \cite{KLA04}
 fits a wide region containing 
several peaks simultaneously after
using a Bayesian procedure to identify the location of the peaks.

The claim for \BBz\ decay was made by a fraction of the 
Heidelberg-Moscow collaboration. A separate group of the
original collaboration presented their analysis of the data at the 
IV International
Conference on Non-Accelerator New Physics\cite{BAK03}. 
They indicate that the data
can be separated into two distinct sets with different 
experimental conditions. One set includes
events that are described as ``underthreshold pulses'' and one set 
that does not. Analysis of the two
sets produce very different conclusions about the presence of the 
claimed peak. They conclude that the evidence
is an experimental artifact and not a result of \BBz decay.  KDHK responds\cite{KLA04} 
that these corrupt data were not 
included in their analysis.

Traditionally, \BB\ experiments have ignored systematic 
uncertainties in their analysis. Only recently with
the start-up of high-statistics \BBt\ results has this situation 
begun to change. Historically, \BBz\ results
have always been quoted as upper limits based on low count rates. 
As a result, systematic uncertainties tended to
be negligible in the final quoted values. With a claim of a 
positive result, however, the stakes are dramatically
raised. It is clear that it is difficult to produce a convincing 
result when the signal counts are comparable
to expected statistical fluctuations in the background. The 
further presence of nearby unidentified peaks
makes the case even harder to prove. Although KDHK does discuss 
some systematic uncertainties qualitatively
and indicates they are small (in the position of the \BBz\ peak, 
and the expected peak width, for example),
there is no consideration of an uncertainty associated with the 
background model.

The next round of proposed \BBz\ experiments are designed to reach
$\sqrt{\delta m_{\rm atm}^{2}}$ and therefore will quickly
confirm or repudiate this claim. This is fortunate since the 
feature near 2039 keV in the KDHK claim
will likely require an experimental test. These experiments should 
provide a detailed listing of all
identified systematic uncertainties and a quantified estimate of 
their size. Furthermore, because the stakes
are very high and there will be many people who are biased, either 
for or against the KDHK claim, blind analyses
should also become part of the experimental design.

\subsection{The Search for Decays to Excited States}
Searches for \BB\ to excited states in the daughter atom have been performed
in a number of isotopes but only observed in $^{100}$Mo (The  experimental
situation is reviewed by Barabash\cite{BAR00}.) and $^{150}$Nd\cite{BAR04}.
These experiments typically search for the $\gamma$ rays that 
characterize the excited states and
therefore are not mode-specific searches. The interpretation 
therefore is that the measured rate
(or limit) is for the \BBt\ mode. These data may be very useful to 
QRPA nuclear theory because
the behavior of the nuclear matrix elements with respect to $g_{pp}$ 
for the excited state decays is different
than for transitions to the ground state\cite{GRI92,AUN96,SUH98}. Thus, the excited state 
transitions probe different aspects
of the theory and may provide insight into the physics of the matrix elements.

A further reason for interest in decays to the excited state, as 
mentioned earlier, is the potential ability to discover
the process mediating the decay\cite{SIM02,TOM00}.
However, the decay rate to an excited state is 10-100 
times
smaller than rate to the ground state\cite{SUH00a,SUH00b}. 
Furthermore the structure of the excited state in the daughter 
nucleus is
not as well understood as the ground state, and this increases the 
relative uncertainty in the nuclear matrix element.

\subsection{The Search for $\beta^+\beta^+$ Modes of Decay}
The $\beta^+$ modes of decay have not received the attention of the 
$\beta^-$ modes because of the
greatly reduced phase space and corresponding long half-lives. 
However, the rate is proportional to $\langle m_{\beta\beta} \rangle^2$
and its detection would
provide additional matrix-element data. Furthermore, if the 
zero-neutrino mode were detected,
it might provide a handle on whether the decay is predominantly
mediated by a light neutrino or by right-handed currents\cite{HIR94}.

Radiative neutrinoless double electron capture is a possible 
alternative to traditional neutrinoless
double beta decay\cite{SUJ03}. In this process, two 
electrons are captured from
the atomic electron cloud and a radiated photon carries the full 
Q-value for the decay. A resonance condition can enhance the rate 
when the energy release
is close to the 2P-1S energy difference. In this case, high-Z, 
low-energy-release isotopes are favored (e.g. $^{112}$Sn).
Unfortunately the mass differences for the candidate isotopes are not 
known precisely enough to
accurately predict the overlap between the two energies. If a 
favorable overlap does exist, however, the
sensitivity to \mee\ might rival that of \BBz decay.

\subsection{Towards a 100-kg experiment}
The KDHK spectrum shows a feature very close to the \BBz\ endpoint. 
This intriguing result will need to be
confirmed or refuted experimentally. One can see the required 
operation parameters for a confirmation experiment
from the KDHK result. One needs about 75 kg-y 
of exposure,
and a background lower than about 0.5 counts/(kg y).  Note that 
most of the proposals described above will all accomplish this very 
early on in their program if they meet their design goals.
If instead one designs an experiment only to test the claim (not to 
provide a precise measurement of the
\Tz) then a 100-kg experiment could provide the answer after a modest run time.

If the KDHK result holds up, it will be a very exciting time for 
neutrino-mass research. A \mee\ near 400 meV
means that $\beta$-decay experiments and cosmology will be sensitive 
to the mass. As a result, one can certainly
imagine a not-too-distant future in which we know the neutrino mass 
and its Majorana-Dirac character. Towards
this goal, a precision measurement of \mee\ will be required. To 
accomplish this, we will need more than
one \BB\ experiment, each with a half-life measurement accurate to 10-20\%.
At this level the uncertainty will
be dominated by the matrix element uncertainty even if future 
calculations can be trusted to 50\%.
With two experiments utilizing different isotopes, one might 
disentangle the uncertainty in \Mz.

\subsection{Towards a 100-ton experiment}
The next generation of experiments hopes to be sensitive to 
$\sqrt{\delta m_{\rm
atm}^{2}}$. If they fail
to see \BBz\ at that level, the target for the succeeding 
generation of efforts will be
$\sqrt{\delta m_{\rm sol}^{2}}$. This scale is an order of magnitude 
lower and hence will require two
orders of magnitude more isotopic mass, 
approximately 100 tons of isotope.

\subsection{Proposed Experiments for \BB\ Decay}
The recent reviews by Elliott and Vogel\cite{ELL02} and Elliott and Engel\cite{ELL04} describe
the basics of 
experimental \BBz\ decay in some detail. Therefore,
we refer the reader to those articles and only summarize the status of 
the various projects.
Table \ref{tab:ZeroNuProj} lists the proposals.

\begin{table}
\caption{\protect  A summary of the \BBz\ proposals. Background 
estimates were not available for all projects. The quantity
of isotope includes the estimated efficiency for \BBz.}
\label{tab:ZeroNuProj}
\begin{center}
\begin{tabular}{lrcl}  \hline\hline
Collaboration                 & Isotope               &     Anticipated & Detector    \\
                              &    (kmol)             & Background & Description   \\
                              &                       &            (counts/y) &     \\ \hline
     CAMEO\cite{BEL01}       &    $^{116}$Cd (2)     &      few/year & CdWO$_{4}$ crystals in liq. scint. \\
     CANDLES\cite{KIS04}      &    $^{48}$Ca (0.04)   & & CaF$_{2}$ crystals in liq. scint.            \\
     COBRA\cite{ZUB01}        &                       & & CdTe semiconductors             \\
     CUORE\cite{ARN04a}       &    $^{130}$Te (1.4)   &    $\approx$ 60/y & TeO$_{2}$ bolometers           \\
     DCBA\cite{ISH00}         &    $^{82}$Se (2)      &    $\approx$ 40/y & Nd foils and tracking chambers             \\
     EXO\cite{DAN00}          &    $^{136}$Xe (4.2)   &    $<1$/y & Xe TPC,              \\
     GEM\cite{ZDE01}          &     $^{76}$Ge (11)    &    $\approx$ 0.8/y & Ge detectors in LN             \\
     GENIUS\cite{KLA01a}       &     $^{76}$Ge (8.8)   &    $\approx$ 0.6/y & Ge detectors in LN             \\
     GSO\cite{DAN01,WAN00}    &     $^{160}$Gd (1.7)  & & Gd$_{2}$SiO$_{5}$ crystals in liq. scint.             \\
     Majorana\cite{GAI03}     &      $^{76}$Ge (3.5)  &    $\approx$ 1/y & Segmented Ge detectors            \\
     MOON\cite{EJI00}         &      $^{100}$Mo (2.5) &    $\approx$ 8/y & Mo foils and plastic scint.             \\
     MPI bare Ge\cite{ABT04}  &     $^{76}$Ge (8.8)   & & Ge detectors in LN             \\
     Nano-crystals\cite{MCD04}& $\approx$ 100 kmol     & & suspended nanoparticles             \\
     Super-NEMO\cite{SAR00}   &     $^{82}$Se (0.6)   &     $\approx$ 1/y & foils with tracking             \\
     Xe\cite{CAC01}           &     $^{136}$Xe (6.3)  &     $\approx$ 118/y & Xe dissolved in liq. scint.             \\
     XMASS\cite{MOR01}        &    $^{136}$Xe (6.1)   & & liquid Xe             \\  \hline
\end{tabular}
\end{center}
\end{table}

\subsubsection{CANDLES}
The CANDLES collaboration has recently published the best limit on 
\BBz\ decay of $1.4 \times 10^{22}$ y
in $^{48}$Ca\cite{OGA04}. Using the ELEGANTS VI detector, this
experiment consisted of 6.66 kg of CaF$_2$(Eu) crystals surrounded by 
CsI crystals, a layer of Cd, a layer of Pb,
a layer of Cu, and a layer of LiH-loaded paraffin, all enclosed 
within an air-tight box. This box
was then surrounded by boron-loaded water tanks and situated 
underground at the Oto Cosmo Observatory.
This measurement successfully demonstrated the use of these crystals 
for \BB\ studies.

An improved version of this crystal technology, the CANDLES-III 
detector\cite{KIS04}, is presently being constructed with 
200 kg
of CaF$_2$ crystals. These crystals have better light transmission 
than the CaF$_2$(Eu) crystals.
This design uses
sixty 10-cm$^3$ CaF$_2$ crystals, which are immersed in liquid
scintillator. The collaboration has
also proposed a 3.2-t experiment that hopes to reach 100 meV for \mee.

\subsubsection{COBRA}
The COBRA experiment\cite{ZUB01} uses CdZnTe or CdTe semiconductor 
crystals. These crystals have many of the advantages of
Ge detectors but, in addition, operate at room temperature. Because 
the crystals contain Cd and Te, there are 7 \BB\ and
$\beta^+\beta^+$ isotopes contained. The final proposed configuration 
is for 64000 1-cm$^3$ crystals for
a total mass of 370 kg. The collaboration
has already obtained 30-keV resolution at 2.6 MeV with these 
detectors and has published initial \BB-decay studies\cite{KIE03}. Background
studies are the current focus of the efforts. Although it is tempting 
to focus on the naturally isotopic abundant
$^{130}$Te for \BBz\ decay, the presence of the
higher Q-value $^{116}$Cd creates a serious background from its \BBt\ 
decay. Detectors enriched in $^{116}$Cd are
probably required to reach 45 meV.

\subsubsection{CUORE}
CUORE and CUORICINO are based on the technique of cryogenic detectors. 
When operated at low temperature, the absorbers of these detectors have 
a heat capacity so low that even the small energy released by a single 
radioactive decay event can be observed and measured by means of a suitable 
thermal sensor. With crystals of mass near to a kilogram, with NTD Ge 
(Neutron Transmutation Doped germanium) thermistors, an 
energy resolution similar to that of germanium diodes has been achieved. In addition, 
thermal detectors allow a wide choice of nuclei to be used for double 
beta decay searches. The experiment 
CUORICINO is located in the Gran Sasso underground laboratory and it is
a prototype for CUORE (Cryogenic Underground 
Observatory for Rare Events). CUORICINO is an array of 44 crystals 
of TeO$_2$ each 5x5x5 cm and 18 crystals each 3x3x6 cm. With its mass 
of approximately 40 kg, CUORICINO is by far the most massive cryogenic
set-up in operation. Due the large isotopic abundance (34\%)
of the double beta decay candidate \nuc{130}{Te}, no isotopic enrichment 
is required, but two of the 3x3x6 cm crystals are enriched in \nuc{130}{Te} 
and two other in \nuc{128}{Te} to investigate \BBt. 
In only three months of operation, CUORICINO has obtained a 90\% c.l.
limit on the lifetime against neutrinoless double beta decay of $7.5 \times 10^{23}$ yr\cite{ARN04},
corresponding to an upper limit on the average neutrino mass ranging from 0.3 
to 1.7 eV.  This result rivals the best limits obtained from many 
years of searches for the double beta decay of $^{76}$Ge. CUORE will consist of an 
array consisting of 25 columns of 10 planes of 4 TeO$_2$ crystals each 5x5x5 cm 
of for a total of 1000 crystals with a mass around 760 kg. Each tower will 
therefore be similar to the single tower of CUORICINO, which consists of 13
planes.  As far as time is concerned, CUORICINO is now running and its larger
brother CUORE will be available in four years from the start of construction 
(likely summer of 2004).  

The present 
background of CUORICINO in the region of neutrinoless double beta decay 
(0.20+0.02 counts/keV/kg/year) is in excellent agreement with the previously 
predicted value (0.22 counts/keV/kg/year). In the present measurement 
this background is mainly due to the surface contamination of copper and 
crystals and recently it has become understood how to reduce it by an order of magnitude 
by surface treatment. Taking into account that the structure of CUORE allows a 
large suppression of background by applying the anticoincidence method we can 
guarantee a conservative value of background of 0.01 counts/keV/kg/year.  We 
believe that in the next four years we can achieve a further improvement in 
the energy resolution, in the radioactive contamination, and in the neutron
and cosmic ray background.  Thus, it is reasonable to predict a background 
of 0.001 counts/keV/kg/year for CUORE.  As a consequence we believe that 
the CUORE sensitivity can be in the few tens of millielectronvolts for 
the average neutrino mass\cite{ARN04a}. 

\subsubsection{DCBA}
As demonstrated by the history of discoveries of extremely rare events,
magnetic tracking detectors have
played important roles. Thus this collaboration bases its 
search for \BBz\ 
events on a magnetic tracking detector. A momentum 
analyzer called the Drift Chamber Beta-ray Analyzer (DCBA)\cite{ISH00} is a tracking 
detector operated in a uniform magnetic field of around 1 kG. 
Various isotopes can be installed within the detector, if 
the source can be fabricated into a thin plate. Presently,
the collaboration is considering \nuc{82}{Se}, \nuc{100}{Mo}, and \nuc{150}{Nd}
because of their high Q-values. 
A tracking region on each side of a source plate includes anode-,
potential- and cathode-wires. The drift region is filled with 1-atm 
helium gas mixed with small amounts of a quench gas. A  $\beta$ ray 
emitted from a source plate makes a helical track in the region 
between anode and cathode wire planes. Anode signals are read 
out with Flash Analog to Digital Converter (FADC). The three-dimensional 
reconstruction of a helical track is available using data from the electron 
drift time (corresponding to the X-coordinate), an anode wire position (Y) 
and the ratio of signals from both sides of an anode wire (Z). Momentum 
of each $\beta$ ray is derived from the curvature of the track. 
Since financial support for DCBA has  
not been approved yet, the construction schedule is unknown. 
Electron tracks of 1 MeV were studied using internal conversion 
electrons from \nuc{207}{Bi}, which was installed in a prototype called DCBA-T. 
We are making efforts to improve the Z-position resolution so as to obtain 
better energy resolution. Another developing item is to accommodate 
source plates, as much as possible, in a limited chamber volume. 
Research into cleaning the source material is also proceeding. 

The future DCBA experiment will consist of 40 modules. One 
module comprises a drift chamber of about 1.8 m$^3$ volume containing the 
source plates surrounded by a solenoid magnet with maximum 
field of 1.6 kG. In the chamber, 30 tracking regions cover 29 source 
plates. The total source-plate area is 25 m$^2$ in each module. For 
a source-plate thickness of 60 mg/cm$^2$, the source weight is 15 
kg for each module. Therefore total source weight is 600 kg, corresponding 
to about 6600 mol and 5400 mol for \nuc{82}{Se} and \nuc{100}{Mo}, respectively, which are
enriched to 90\%.  For a natural Nd source, about 200 mol of \nuc{150}{Nd} will be 
installed. Assuming an efficiency of event detection of 
0.3, a background rate of 1 event/module/year, and a measuring time 
of 5 years;  the half-life sensitivities are approximately calculated to
be $3 \times 10^{26}$ yr for \nuc{82}{Se}, $2 \times 10^{26}$ yr
for \nuc{100}{Mo} and $9 \times 10^{24}$ yr for \nuc{150}{Nd} in 
natural Nd. If 90\% enriched \nuc{150}{Nd} is available in the future by the method 
of Atomic Vapor Laser Isotope Separation, it is possible to obtain $1 \times 10^{26}$ 
yr for \nuc{150}{Nd}.

\subsubsection{EXO}
The {\bf E}nriched {\bf X}enon {\bf O}bservatory\cite{DAN00} is being set-up to study
the double beta decay of $^{136}$Xe. EXO is a collaborative effort 
currently involving groups from Caltech, Carleton University (Canada), 
Colorado State University, ITEP Moscow (Russia), Laurentian University (Canada),
SLAC, Stanford University, University of Alabama, University of California Irvine,
Universite de Montreal (Canada), Universite de Neuchatel and Universite de Yverdon
(Switzerland).
The collaboration aims to use up to 10 tons of isotopically enriched $^{136}$Xe
to build a redundant and background-free detector using good energy resolution, 
pattern recognition and the identification of the atomic species produced by the 
double-beta decay.   A concise description of the project has been published 
in~\cite{DAN00}.

The final state atom tagging is possible because of the simple and well known
atomic spectroscopy of Ba$^+$ ions.   Such spectroscopy has enabled the observation
of individual ions illuminated with appropriate wavelengths since about 20 years.
The specific wavelengths needed to produce atomic fluorescence ensure extreme
selectivity of this technique.     Ba happens to be the atomic species produced in
the double beta decay of $^{136}$Xe.    In EXO the Xenon will be used as an active 
target in a Time Projection Chamber (TPC) either in liquid (LXe) or gas (GXe) phase. 
In the GXe case the laser beams would be steered to the location where a candidate
decay has occurred.   In the LXe case the Ba-ion candidate would be extracted and 
brought into an ion trap where the fluorescence would be observed.    The possibility
of observing the fluorescence of the Ba directly in the liquid is also being investigated
by one of the EXO groups.    While R\&D is proceeding at different institutions
for both liquid and gas phase TPC, a LXe TPC for 200~kg of $^{136}$Xe is being built 
as a prototype and as a first step towards the very large detector.

EXO plans to test Majorana neutrino masses as small as 10 to 40 meV using this 
scheme~\cite{DAN00} that should ensure extremely high background rejection power. 
This sensitivity covers neutrino masses derived from the atmospheric mass spitting. 
A degenerate or inversely hierarchical neutrino mass pattern can thus be tested with EXO.

$^{136}$Xe is a particularly convenient isotope for a very large double beta decay experiment.
It combines a large Q-value with ease (and low cost) enrichment, the absence of long lived
radioactive isotopes, ease of purification and the possibility of transfer from one detector
to another in the case new technology would become available.    In addition purification 
can be achieved on-line so that more refined purification system may be introduced as they
become needed.    Finally, together with the enabling possibility of final state 
identification that is the hallmark of EXO, $^{136}$Xe has the longest 
\BBt\-half life among all high Q-value $\beta\beta$-unstable nuclides 
(at least factor 6.5 longer than e.g. $^{76}$Ge).

With its 200 kg of source strength the EXO prototype detector will 
already represent the largest existing double beta decay experiment. Recently 
evidence for a Majorana neutrino mass of $0.44^{+0.14}_{-0.20}$ eV (errors 
given at 3$\sigma$ c.l.) has been claimed by a part of the Heidelberg component 
of the Heidelberg-Moscow experiment~\cite{KLA04}.    Should this claim
turn out to be correct the EXO prototype expects to observe 
$43^{+33}_{-30}$ \BBz\ decays per year when using the same matrix 
element calculation as reference~\cite{KLA04}.
Assuming that funding continues to be available in a timely fashion, the prototype 
detector is expected to be commissioned at Stanford in the Winter 2004-5 and be 
transferred to WIPP in the Summer 2005.
The development of the barium extraction and tagging will continue in parallel,
together with the work on a GXe TPC.   After initial operation of the prototype
the technology for the large detector will be chosen and its design finalized.

\subsubsection{MOON}
MOON (Molybdenum Observatory Of Neutrinos)\cite{EJI00} is a "hybrid" \BB\ and solar \nue\ 
experiment with \nuc{100}{Mo}. It aims at studies of \mee\ with a 
sensitivity near 30 meV by measuring \BBz\ decays of 
\nuc{100}{Mo} and the charged current \nuc{7}{Be} solar \nue\ with an accuracy of 
about 10\% by inverse $\beta$ decays of \nuc{100}{Mo}.
The \BB\ decays to the ground and excited states are measured in prompt 
coincidence for the \BBz\ studies. The large Q=3.034 MeV results in a large 
phase-space factor to enhance the \BBz\ rate and a large energy to 
place the \BBz\ signal above most background. 
MOON is a spectroscopic study of two $\beta$ rays. As such, its capability to measure
the energy and angular 
correlations for the two $\beta$ rays can help identify the \BBz\ mechanism. 
It capability for a tight localization of  \BB\ events in space and time is crucial 
for selecting \BBz\ and reducing background. 
 
A possible configuration of the MOON apparatus is a super module of hybrid plate and 
fiber scintillators with $^{100}$Mo isotope mass totaling about 1 ton. One module 
of this apparatus
consists of a thin (20 mg/cm$^2$) Mo film interleaved between X-Y fiber planes
and a plate scintillator. The fiber scintillators enable one to get the 
position resolution (1/K $< 10^{-8}$ pixels per ton) and the scintillator plate provides 
an energy resolution ($\sigma \approx$2.2\% at 3 MeV). A different detector option under 
consideration consists of foils within liquid Ar to obtain better energy resolution. 
 Research has shown (1) that for the small plate scintillator, 
 $\sigma$ = 2.2\% , FWHM: 5\% for \BBz\, including light attenuation,
 for MOON, (2) the position resolution of $3.2 \times 10^{Ð9}$, and (3) 
 the feasibility of using centrifugal separation of MoF$_6$ gas to produce
 $^{100}$Mo enrichment of 85-90\% in ton-scale quantities. Currently a proto-type MOON
 (MOON 1) is under construction (2003-2005). A proposal for MOON is planned by 2006. 

  The MOON sensitivity can be evaluated as follows. The source is 1 ton of
  Mo with 85\%$^{100}$Mo. The \BBz\ efficiency after energy and angle cuts is 0.28. 
  The background from \BBt\ in the \BBz\  window during 5 years after cuts is 5.5(42) events
  for an energy resolution of $\sigma$=2.2(3)\% (FWHM 5(7)\%). Here $\sigma$=2.2\% is estimated
  based on the R\&D data from a small-scale prototype and ELEGANT V, while 3\%
  is a conservative value including a 50\% factor. The background from cosmogenic isotopes
  and natural isotopes are less than 0.5 events. Then the \BBz\ yield of $\sqrt{background}$
  is 2.3(6.4), 
  corresponding to a half life 22 (7.7 )$\times 10^{26}$. The mass sensitivities are, respectively,
  13 (22) meV. Here the matrix element of \Mz\ = 3 is estimated by 
  referring to the recent 3 calculations.  
The MOON apparatus has  the \BB source separated from the detector, and therefore 
can be used for other 
\BB\ isotopes such as \nuc{82}{Se}, \nuc{150}{Nd} and \nuc{116}{Cd}
as well by replacing Mo isotopes with 
other isotopes.

\subsubsection{Majorana}
The Majorana Collaboration proposes to field 500 kg of 86\% enriched 
Ge detectors\cite{GAI03}.
By using segmented crystals and pulse-shape analysis, multiple-site 
events can be identified
and removed from the data stream. Internal backgrounds from 
cosmogenic radioactivities will be greatly
reduced by these cuts and external $\gamma$-ray backgrounds will also 
be preferentially eliminated.
Remaining will be single-site events like that due to \BB. The 
sensitivity is anticipated to be $4 \times
10^{27}$ y. 

Several research and development activities are currently proceeding. 
The collaboration is building a multiple-Ge detector array, referred 
to as MEGA,
that will operate underground at the Waste Isolation Pilot Plant 
(WIPP) near Carlsbad, NM USA. This experiment will investigate the 
cryogenic cooling of many detectors sharing a cryostat in addition
to permitting studies of detector-to-detector coincidence techniques 
for background and signal identification.
A number of segmented crystals are also being studied to understand 
the impact of segmentation on
background and signal. Recently, these segmented detectors have shown
that the radial information provided by PSD is indeed orthogonal to 
the azimuthal information provided by the segmentation.
The segmentation studies, or SEGA program, consists of one 12-segment 
enriched detector and a number of commercially
available segmented detectors. Presently, commercially available 
segmented detectors are fabricated from
n-type crystals. Such crystals are much more prone to surface damage 
and thus more difficult to handle when
packaging inside their low-background cryostats. Hence the 
collaboration is also experimenting with segmenting p-type
detectors.

The Majorana design uses Ge detectors within a low-mass, electroformed
Cu cryostat. Electroformed Cu is very free of radioactive contaminants.
However, just how radio-pure the Cu is remains unknown. Hence, the collaboration 
plans to form Cu underground and study its radiopurity to a
sensitivity below previous limits.

\subsubsection{Bare Ge Crystals}
The GENIUS collaboration\cite{KLA01a} proposed 
to install 1 t of enriched
bare Ge crystals in liquid nitrogen. By eliminating much of the 
support material surrounding
the crystals in previous experiments, this design is intended to 
reduce backgrounds of external origin.
Note how this differs from the background-reduction philosophy 
associated with pulse-shape analysis coupled
with crystal segmentation.
The primary advocates for this project indicate\cite{KLA04} 
that its motivation has
been questioned by their own claim of evidence for \BBz decay. Even 
so, the GENIUS test facility\cite{KLA03}
is being operated to demonstrate the effectiveness of operating 
crystals naked in liquid cryogen.

Another group at the Max Plank Institute in Heidelberg, however, is 
proposing to pursue a similar idea.
They have recently submitted a Letter of Intent\cite{ABT04} to the 
Gran Sasso Laboratory. They propose
to collect the enriched Ge crystals from both the Heidelberg-Moscow 
and IGEX expeiments and
operate them in either liquid nitrogen or liquid argon. As a second 
phase of the proposal, they
plan to purchase an additional 20-kg of enriched Ge detectors (most 
likely segmented) and operate with a total of 35 kg
for about 3 years. Finally, they eventually plan to propose a large 
ton-scale experiment.
It should be noted that this collaboration and the Majorana 
collaboration are cooperating on
technical developments and if a future ton-scale experiment using 
$^{76}$Ge proceeds these
two groups will most likely merge and optimally combine the 
complementary technologies of bare-crystal operation
and PSA-segmentation.

\subsubsection{Nanocrystals}
Some elements may be suitable for
loading liquid scintillator with metallic-oxide nanoparticles. Since
Rayleigh scattering varies as the sixth power of the particle radius, it
can be made relatively small for nanoparticles of radii below 5 nanometers.
Particles of this size have been developed
and commercial suppliers of ZrO$_{2}$, Nd$_{2}$O$_{3}$ {\it etc}. are 
available.
Absorption of the materials must also be taken into account, but some of
the metal oxides such as ZrO$_{2}$ and TeO$_{2}$ are quite 
transparent in the optical
region because of the substantial
band gaps in these insulators. Some members of the SNO collaboration\cite{MCD04} have been studying
a configuration equivalent to filling the SNO cavity with a 1\% 
loaded liquid scintillator or approximately 10 t
of isotope after the present heavy water experiment is completed.
The group is currently researching the optical properties of
potential nano-crystal solutions. In particular, one must demonstrate 
that sufficient
energy resolution is achievable with liquid scintillator.

\subsubsection{Super-NEMO}
The currently operating NEMO-3 detector uses a tracking-calorimeter 
technique to detect \BB. The source, is not the detector, but rather 
a thin foil located in the middle of a tracking chamber. This 
tracking chamber, which is made of geiger-drift cells, is surrounded 
by a calorimeter (scintillators with low radioactivity PMTs). The 
first of two long runs to search for \BBz\ decay began in February 
of 2003 with 7 kg of \nuc{100}{Mo} and 1kg of  \nuc{82}{Se}. Additionally, 2 kg of
various foils were placed in the detector to study \BBt\ decay and 
backgrounds. The first run will last approximately five years. So 
far a comprehensive study of the effects of various shields has 
been undertaken.  The second run, again for five years, is 
currently planned to operate with 20 kg of \nuc{82}{Se}. The \nuc{82}{Se} is
of particular interest because the \BBt\ decay lifetime is 10 times 
that of \nuc{100}{Mo} and thus may contribute less as a background to \BBz\ decay.

There has been an unexpected background from radon, which is 
currently being fixed with a  Òradon tentÓ and a "radon trapping factory".
Data collection 
free of the radon background will start in Fall 2004. The energy 
resolution is 250-keV FWHM at 3 MeV. Assuming the radon is eliminated 
with the radon tent and an efficiency
of 14\%, the background will be 5x10$^{-4}$ counts/keV/kg/yr. This yields \Tz\ $> 5 
\times 10^{24}$ in five 
years with \mee\ $ < 200-500$ meV. During the second run with 20 kg of $^{82}$Se,
an efficiency of 14\%
and a background of 5x10$^{-5}$ counts/keV/kg/yr, the expected results in 
five years should be  \Tz\ $> 3 \times 10^{25}$ yr and \mee\ $< 100-200$ meV.

The recent progress of the NEMO-3 program\cite{SAR00} has 
culminated in excellent \BBt\ results.
In particular, the energy spectra from $^{100}$Mo
contain approximately 10$^5$ events and are nearly background free. These 
data permit, for the first
time, a precise study of the spectra. In fact, there is hope that 
the data (Sutton 2004) will demonstrate
whether the \BBt\ transition is primarily through a single 
intermediate state or through a
number of states\cite{SIM01}. 

A much bigger project is currently being planned that would use 100 
kg of source. The apparatus would have a large
footprint however and the Frejus tunnel where NEMO-3 is housed would 
not be large enough to contain it. Currently the collaboration
is studying the design of such a detector.

\subsubsection{XMASS}
The XMASS collaboration\cite{SUZ04,MOR01} plans to build a 
10 t natural Xe liquid scintillation
detector. They expect an energy resolution of 3\% at 1 MeV and hope 
to reach a value for \Tz\ $> 3.3 \times
10^{27}$ y. This detector would also be used for solar-neutrino 
studies and a search for dark matter.

\subsubsection{Borexino CTF}
In August of 2002, operations at the Borexino experiment resulted in 
the spill of scintillator. This
led to the temporary closure of Hall C in the Gran Sasso Laboratory 
and a significant change in operations at the underground
laboratory. As a result, efforts to convert the Counting Test 
Facility (CTF) or Borexino itself into a \BBz\ experiment\cite{BEL01,CAC01}
have been suspended\cite{GIA04}.

\subsection{How many \BB\ Experiments are Required?}

In view of the importance and scale of new generation \BBz experiments, internationally 
cooperative efforts in both experiment and theory are quite important. Thus it is
 reasonable to suggest a concentration of the limited  resources on
 a few  \BB experiments However, it is critical to use different experimental techniques 
and isotopes to demonstrate the effect has really been seen and also to extract the most
critical physics conclusions. It is therefore necessary to build a number of detectors and 
the reasons are enumerated here.

\begin{enumerate}
\item The observation of a statistically significant signal in a single experiment 
might not be considered a discovery without clear confirmation from other
independent experiments utilizing different isotopes.
\item A nuclear matrix element is necessary to deduce  \mee from 
  a measured \BBz rate.  Since theoretical calculations of \Mz may include 
 a substantial uncertainty, one needs experiments on different isotopes to extract a
 reliable value for the effective mass.
\item Although light-neutrino exchange is the most natural explaination
for \BBz if it exists, there are other possibilities. The relative matrix element values for
different nuclei depend on the mechanism. Furthermore the matrix element situation is encouraging and 
one can anticipate a great improvement
in the calculation precision. Therefore, measurements in several nuclei might be the
most straight-forward way to provide insight
into the mechanism of \BBz. 
\item There are a number of different techniques being proposed for future experiments.
 Each has been previously used as effective prototypes for the proposals
and therefore remains a strong candidate for
future effort. However, the sensitivity of any given proposal depends strongly on its  
background estimate. It remains to be seen which of the technologies will 
successfully attain the required
background. In addition, certain technologies
provide capabilities such as measurments of the opening angle, individual electron energies, or 
the daughter production. These will not only help understand and remove background
but they may also provide insight into the mechanism of \BBz.
\end{enumerate}

\section{VII. Conclusions}
Study of the neutrinoless double beta decay and searches for the manifestation
of the neutrino mass in ordinary beta decay are the main sources of information
about the absolute neutrino mass scale, and the only practical source
of information about the charge conjugation properties of the neutrinos.
Thus, these studies have a unique role in the plans for better understanding
of the whole fast expanding field of neutrino physics.

In this report we summarized the various aspects of the problem. We explained
first the relation of the information that can be obtained from the analysis
of \BBz\ and $\beta$ decay experiments with the parameters describing
neutrino oscillations and the absolute neutrino mass scale. We then
discussed the nuclear structure issue and the uncertainty  in determining
the neutrino mass from a measured \BBz\ decay rate. We also briefly
discussed the role of neutrino mass in cosmology, and the corresponding
constraints on neutrino masses based on astrophysical observations.

The remaining part of the report was devoted to the description of the
existing and planned experiments. We described the present situation
in \BBz\ and tritium beta decay, and in particular the recent, so far
unverified, claim of the \BBz\ decay discovery. We concentrated then
on the future plans for both areas of research. 

The situation in the direct neutrino mass searches using tritium 
$\beta$ decay is relatively simple. There is only one realistic plan  
a very large new beta spectrometer which is being built in Germany. 
This KATRIN experiment
has a design sensitivity approaching 200 meV. 
If the neutrino masses are quasi-degenerate, as would
be the case if the recent double-beta decay claim proves true, 
KATRIN will see the effect. Although 
KATRIN is predominately a European effort, there
is significant US participation.
The design and construction of this experiment is proceeding well and
we enthusiastically recommend the
continuing strongly support of this program.

There are many proposals, and even more ideas, for much larger \BBz\
decay experiments than the existing ones. We outlined and justified
in the report
the strategy that we believe should be followed:
\begin{enumerate}
\item A substantial number (preferably more
than two) of 200-kg scale experiments (providing the capability to
make a precision measurement
at the quasi-degenerate mass scale) with large US participation
should be supported as soon as possible.
We estimate that the timescale of such experiments is 3-5 years,
and  that each such experiment will cost approximately \$10M-\$20M.
\item Concurrently, the development
toward $\sim$1-ton experiments ({\it i.e.} sensitive to $\sqrt{\Delta
m_{\rm atm}^{2}}$)
should be supported, primarily as expansions of the 200-kg experiments.
The corresponding plans for the procurement
of the enriched isotopes, as well as for the development of a suitable
underground facility, should be carried out. The US funding agencies
should set up in a timely manner a mechanism to review and compare
the various proposals for such experiments which span
research supported by the High Energy and Nuclear Physics offices of DOE
as well as by NSF. 
Each such experiment will cost approximately \$50M-\$100M and
take 5-10 years to implement.
\item A diverse R\&D program developing
additional techniques of \BBz\ decay study should be supported.
\end{enumerate}

The field of neutrino physics has made great strides recently. We believe
that the study of \BBz\ decay in particular could very well be the next
one where significant discoveries will be made. 
\section{VIII. Acknowledgements}
This working group's co-chairs are Steven Elliott (elliotts@lanl.gov)and 
Petr Vogel(pxv@caltech.edu). The 
members of the working group are
Craig Aalseth (Craig.Aalseth@pnl.gov),
Henning Back (hback@vt.edu),
Loretta Dauwe (dauwe@umich.edu),
David Dean (deandj@ornl.gov),
Guido Drexlin (guido.drexlin@ik.fzk.de),
Yuri Efremenko (efremenk@larry.cas.utk.edu),
Hiro Ejiri (ejiri@rcnp.osaka-u.ac.jp, ejiri@spring8.or.jp),
Jon Engel (engelj@physics.unc.edu),
Brian Fujikawa (bkfujikawa@lbl.gov),
Reyco Henning (RHenning@lbl.gov),
Jerry Hoffmann (hoffmann@physics.utexas.edu),
Karol Lang (lang@mail.hep.utexas.edu),
Kevin Lesko (ktlesko@lbl.gov),
Tadafumi Kishimoto (kisimoto@km.phys.sci.osaka-u.ac.jp),
Harry Miley (harry.miley@pnl.gov),
Rick Norman (norman11@llnl.gov),
Silvia Pascoli (pascoli@physics.ucla.edu),
Serguey Petcov (petcov@he.sissa.it),
Andreas Piepke (andreas@bama.ua.edu),
Werner Rodejohann (werner@sissa.it),
David Saltzberg (saltzbrg@physics.ucla.edu),
Sean Sutton (ssutton@MtHolyoke.edu),
Ray Warner (ray.warner@pnl.gov),
John Wilkerson (jfw@u.washington.edu),
and Lincoln Wolfenstein (lincolnw@heps.phys.cmu.edu).

\end{document}